\documentclass[a4paper,12pt]{article}
\usepackage{graphicx}
\usepackage{bm}
\usepackage{color}
\usepackage{amsmath,amssymb,amsthm,color,CJK,mathrsfs}
\usepackage{multirow}
\usepackage[titletoc]{appendix}
\usepackage{color, colortbl}
\usepackage{natbib}
\usepackage{hyperref, enumitem,authblk}
\usepackage{algorithm}
\usepackage{algpseudocode}
\usepackage[margin=1in]{geometry}

\hypersetup{
    colorlinks=true,
    citecolor=blue
    }

\def\spacingset#1{\renewcommand{\baselinestretch}%
{#1}\small\normalsize} \spacingset{1}
\usepackage[doublespacing]{setspace}
\title{Spatially Penalised Registration of Multivariate Functional Data}
 \author[1]{Xiaohan Guo}   
 \author[1]{Sebastian Kurtek}
 \author[2]{Karthik Bharath}
 \affil[1]{\small Department of Statistics, The Ohio State University, USA}
\affil[2]{\small School of Mathematical Sciences, University of Nottingham, UK}
\date{}
\begin{document}

\maketitle

\begin{abstract}
\spacingset{1} 
\noindent Registration of multivariate functional data involves handling of both cross-component and cross-observation phase variations. Allowing for the two phase variations to be modelled as general diffeomorphic time warpings, in this work we focus on the hitherto unconsidered setting where phase variation of the component functions are spatially correlated. 
We propose an algorithm to optimize a metric-based objective function for registration with a novel penalty term that incorporates the spatial correlation between the component phase variations through a kriging estimate of an appropriate phase random field. The penalty term encourages the overall phase at a particular location to be similar to the spatially weighted average phase in its neighbourhood, and thus engenders a regularization that prevents over-alignment. Utility of the registration method, and its superior performance compared to methods that fail to account for the spatial correlation, is demonstrated through performance on simulated examples and two multivariate functional datasets pertaining to EEG signals and ozone concentrations. The generality of the framework opens up the possibility for extension to settings involving different forms of correlation between the component functions and their phases. 
\end{abstract}

\newpage

\section{Introduction}

Modern functional datasets, such as longitudinal records, medical imaging signals, geometric shapes, contain confounded amplitude and phase variations \citep{srivastava2011registration}. The adverse effects of ignoring the phase variation during their analyses are well-understood \citep{marron2015functional}. The process of registration, also commonly referred to as alignment or amplitude-phase separation, enables the practitioner to decouple the two sources of variation and account for them in further analyses. 
For multivariate functional data $\mathbb R \supset I \mapsto F_i=(f_{i1},\ldots,f_{iK})^\top \in \mathbb R^K$ consisting of multiple correlated, univariate functional components $f_{ij}:I \to \mathbb R,\ j = 1,\ldots,K$, the notion of phase variation can be decomposed into two types: (i) \emph{cross-observation phase}, which is common across all $K$ component univariate functions $f_{ij}$ within an observation $F_i$, and (ii) \emph{cross-component phase} within each observed unit.  

Registration procedures for multivariate functional data  are driven by assumptions on the two types of temporal variation. At the two extremes are procedures which assume that cross-component variation within each $F_i$ are either uncorrelated or identical. The former case is quite commonly considered by researchers in neuroimaging \citep{makeig2007prospects, tsai2014cortical,zhao2020modeling}, motivated mainly by the simplicity of independent componentwise registration. The latter case, referred to as universal registration, is common in shape analysis of $K$-dimensional curves $\{F_i\}$ \citep{kurtek2012statistical}, wherein one assumes that cross-component phase variation does not exist and a common warping function is estimated for all components within the same observation \citep{olsen2016simultaneous}. Thus, neither of the two extreme cases directly addresses the registration problem when cross-component variation is non-negligible. 

Literature on registration methods for multivariate functional data that lie between the two extreme cases is sparse. Noteworthy exceptions within the statistics literature are recent work by \cite{carroll2020cross}, \cite{carroll2021} and \cite{park2017clustering}, where specific forms of cross-component temporal variation were considered. Specifically, the first two works restrict attention to the situation where \emph{all} component functions $f_{ij}$ for $j=1,\ldots,K$ are assumed to have the same shape:  for each observation $i$, \cite{carroll2020cross} assume that time-warped components $f_{ij}(t)$ arise through time shifts $t \mapsto (t -\tau)$ of the same function, say $g$, while \cite{carroll2021} extend this to the case where the time shift map is replaced by a general diffeomorphic warping $t \mapsto \gamma(t)$ of $I$. Motivated mainly by clustering, \cite{park2017clustering} proposed a conditional observation-specific registration procedure to extract relevant features for clustering. 

\begin{figure}[!t]
	\centering
	\includegraphics[scale=0.16]{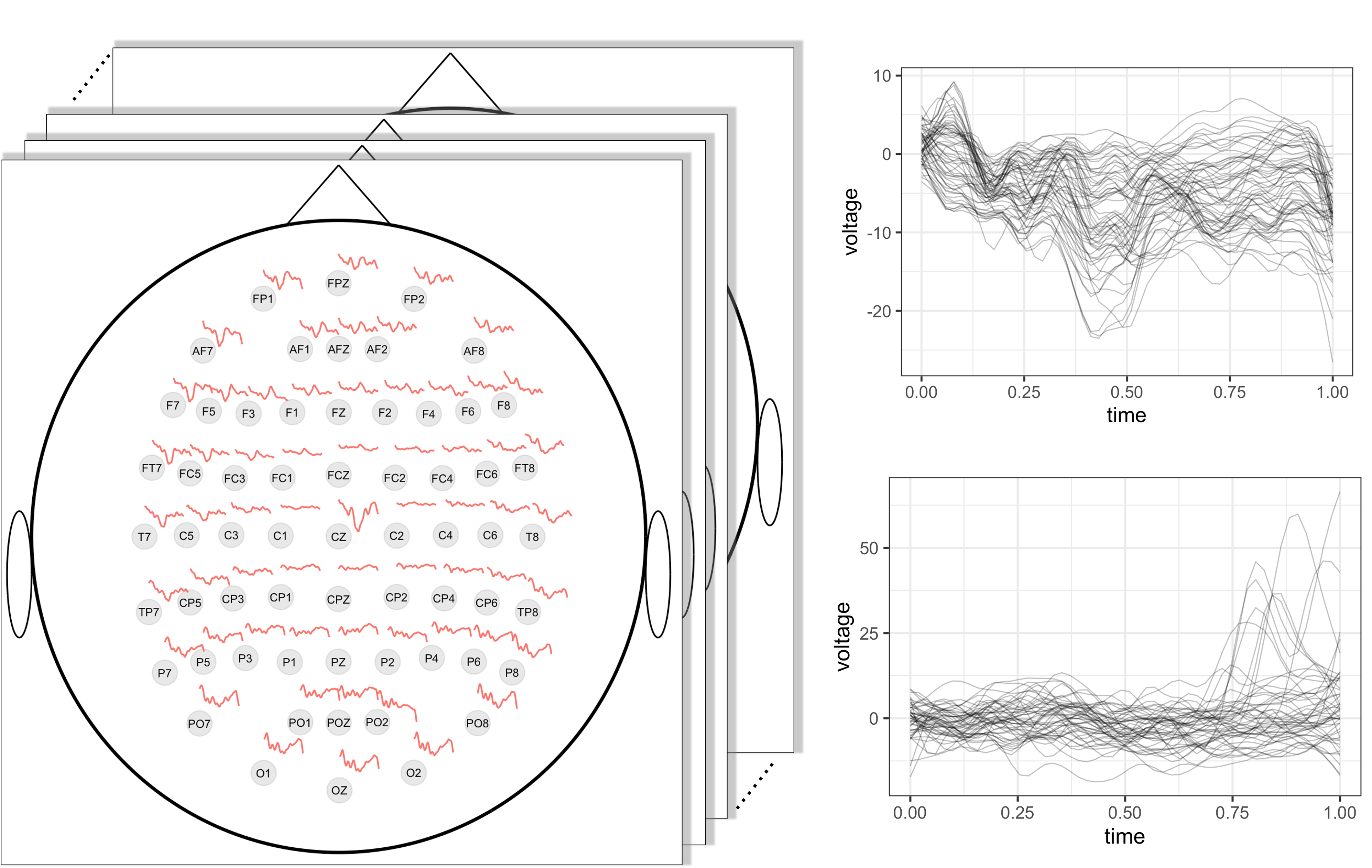}
	\caption{\small Left: Example of a multi-trial EEG dataset with 61 electrodes from an alcoholism study. Each slide represents a single trial in the study, while the red functions plotted at the locations of electrodes (grey circles) constitute the components. Top Right: 61 signals from a single trial with considerable cross-component phase variation. Bottom Right: Signals collected at electrode AFZ across 50 trials with considerable cross-observation phase variation.}
	\label{fig:eeg_example}
\end{figure}

It is quite common to encounter multivariate functional datasets wherein the component functions do not possess the same shape and/or when correlations between component functions need to be explicitly incorporated into the registration procedure. As an example, consider data generated from electroencephalogram (EEG) signals, specifically pertaining to event-related potentials measured during EEG tests. Event-related potentials are small voltages that reflect the electrical activity on the scalp in response to specific stimuli received by study participants \citep{sur2009event}. For the $i$th observation, EEG signals $f_{ij}$ are densely recorded on a certain time domain $I$ using a set of electrodes placed at $K$ different locations on the scalp, and together comprise a multivariate functional observation $F_i$. 
In the left panel of Figure \ref{fig:eeg_example}, we display the observed EEG signals at the (projected) two-dimensional electrode locations on a toy map of the scalp. The entire collection of signals recorded during a single EEG trial is shown as a single slide and corresponds to an observed multivariate functional data unit. 

Evidently, the component functions $f_{ij}$ and $f_{ik}$ at electrode locations $j$ and $k$ during the $i$th trial are spatially correlated due to functional connectivity between brain regions. Latency of the brain’s responses to stimuli varies across trials for a single subject (and among different subjects) resulting in cross-observation phase variation \citep{wang2001warp}. At the same time, different response lags of different brain regions to the presented stimuli \citep{stam2007phase} result in cross-component phase variation. The right two panels of Figure \ref{fig:eeg_example} offer an illustration of cross-observation (bottom) and cross-component (top) phase variation among EEG signals. To account for the two different types of phase variation in multi-trial EEG, an appropriate registration procedure that accounts for spatially-correlated cross-component phase is required. Specifically, the procedure should enable simultaneous registration of all components $f_{ij}$, while taking into account the dependence structure between the cross-component phases. This cannot be achieved by implementing \emph{independent}  registration of $f_{1j},\ldots,f_{nj}$ for each component $j$, since such a procedure results in additional cross-component phase variation in the estimated average components, since each component is treated independently. We note this phenomenon in the right panel of Figure \ref{fig:alcohol_variogram} (and the associated discussion) in Section \ref{sec:EEGanalysis} during our detailed analysis of the EEG data.


Motivated by the type of registration task associated with the EEG data, we propose a registration procedure that exploits correlated phase variation between component functions $\{f_{ij}\}$ of multivariate functional data $\{F_i\}$ and offers a compromise between the two extreme cases of independent componentwise and universal registration methods alluded to above.  Our focus is on multivariate functional data with  component functions $f_{ij}$ whose overall phase variation, represented through a time-warping function $\gamma_{ij}:I \to I$, is a combination $\gamma_{ij}=\alpha_i \circ \xi_j$ of cross-observation phase $\alpha_i:I \to I$ and \emph{spatially correlated} cross-component phase $\xi_j:I \to I$ ($\circ$ is function composition). The flexibility afforded, and desired, by such a general phase specification while registering multivariate functional data with correlated components, such as the EEG data, comes at a cost: the individual phase components $\alpha_i$ and $\xi_j$ cannot be decoupled and estimated individually, and incorporating spatial information present in the latent $\{\xi_j\}$ when estimating $\gamma_{ij}$ is hugely challenging. 
In view of this, our main contributions are as follows.
\begin{enumerate}[label=(\roman{*}), leftmargin=*]
\itemsep 0em
\item We propose a metric-based penalized registration method for multivariate functional data with a penalty term defined using the spatial correlations between the cross-component phases $\{\xi_j\}$, using which the overall phases $\{\gamma_{ij}\}$ are estimated  by eliminating cross-observation phase variation $\{\alpha_i\}$. 
\item The penalty term is defined using a spatially weighted combination of the (estimated) cross-component phases $\{\xi_j\}$, and is, owing to the invariant property of the metric to (simultaneous) time warping, impervious to the cross-observation phases $\{\alpha_i\}$; this allows us to entirely circumvent having to estimate the $\{\alpha_i\}$. 
\item With respect to cross-observation registration, we demonstrate clear superiority of our method  over ones that fail to account for the spatial correlation under a variety of simulation settings and two real data examples. 
\end{enumerate} 
The rest of this paper is organized as follows. Section \ref{setup} details the registration problem for multivariate functional data whose component functions have spatially correlated phase variation. Section \ref{sec:univfundata} briefly reviews the geometric elastic functional data analysis framework, including univariate pairwise and multiple registration, and registration of multivariate functions with common cross-component phase. Section \ref{sec:proposedmethod} details the proposed spatially penalised registration objective function that accounts for spatial correlation in cross-component phase, and provides an algorithm (Algorithm \ref{alg:jointregist}) for its optimization. Section \ref{sec:simu} presents results of simulation studies; in Section \ref{sec:realdata}, we apply the proposed registration approach in two real data scenarios.

\section{The registration problem}
\label{setup}
For a positive integer $r$, denote by $[r]$ the set $\{1,\ldots,r\}$.  The notation $j \in [r]$ will be used in place of $j=1,\ldots,r$.  We can view multivariate functional data $\{F_1,\ldots,F_n\}$ with $F_i=(f_{i1},\ldots,f_{iK})^\top,\ i \in [n]$ with cross-observation and correlated cross-component temporal variation as being generated from the model
\begin{equation}
\label{eq:model}
F_i \circ \bm w_i (t)=\mu(t) +E_i(t), \quad t \in [0,1],\ i \in [n],
\end{equation}
where $\mu=(\mu_1,\ldots,\mu_K)$, with $\mu_j:[0,1] \to \mathbb R$ , is a deterministic mean/template multivariate function, $E_i=(e_{i1},\ldots,e_{iK})^\top$ are realisations of an error process with spatially correlated component functions $\{e_{ij}:[0,1] \to \mathbb R\}$, and $\bm w_i=(\gamma_{i1},\ldots, \gamma_{iK})$ is a vector of monotone increasing, end-points preserving, time warping functions assuming values in $\Gamma:=\{\gamma:[0,1]\to[0,1]\mid\gamma(0)=0,\gamma(1)=1,\dot\gamma>0\}$, where  $\dot\gamma$ is the time derivative of $\gamma$; here, $F_i \circ \bm w_i (t)=(f_{i1}\circ \gamma_{i1},\ldots,f_{iK}\circ \gamma_{iK})^\top$ represents componentwise warping, where $\circ$ denotes function composition.  Note that $\Gamma$ is a group with operation $\circ$, identity element $\gamma_{id}(t)=t$, and the function inverse as the group inverse.  We will interchangeably refer to $\gamma$ as a warping function or phase depending on context. 

The characterising feature of the problem is that, for $ i \in [n],\ j \in [K]$, the warping functions $\gamma_{ij}=\xi_{\mathbf s_j} \circ \alpha_i$, where $\xi_{\mathbf s_1},\ldots,\xi_{\mathbf s_K}$ are spatially correlated warping functions at $K$ spatial locations $\mathbf s_1,\ldots, \mathbf s_K$ within a spatial domain $\mathcal D \subset \mathbb R^p$, with $\xi_{\mathbf s_j},\ \alpha_i \in \Gamma$ for each $i \in [n],\ j \in [K]$. The phase $\gamma_{ij}$ of a component function $f_{ij}$ is thus represented as a combination of warping functions corresponding to cross-observation and cross-component temporal variations. The index $j$ will be used to denote the spatial location $\mathbf s_j$, i.e., $\xi_j$ will be used in place of $\xi_{\mathbf s_j}$, and the two notations will be used interchangeably. 

Under the above setup, we can interpret $\alpha_1,\ldots,\alpha_n$ as representing cross-observation temporal variation within $\{F_i\}$ and $\xi_1,\ldots,\xi_K$ as representing spatially correlated cross-component temporal variation within $\{F_i\}$. The two-fold objective is to,  jointly:
\begin{enumerate}
	\itemsep 0em
	\item register $F_1,\ldots,F_n$ by estimating $ \{\gamma_{ij}: i \in [n],\ j \in [K]\}$ in a manner that ensures for every $i \in [n]$, the warping functions $\gamma_{ij}$ and $\gamma_{ik}$ in nearby spatial locations $\mathbf s_j$ and $\mathbf s_k$ in $\mathcal D$ are `similar', owing to `similar' cross-component warpings $\xi_j$ and $\xi_k$;
	\item estimate components $\mu_1,\ldots,\mu_K$ of the template $\mu$. 
\end{enumerate}
In the context of the EEG application, e.g., in the $i$th trial for a single subject, we desire signals $f_{ij}$ and $f_{ik}$ obtained from spatially proximate electrode locations $\mathbf s_j$ and $\mathbf s_k$ to have similar phase variation $\gamma_{ij}$ and $\gamma_{ik}$, with respect to template components $\mu_j$ and $\mu_k$. The requirement is compatible with findings by \cite{stam2007phase} on different response lags of spatially disparate brain regions to presented stimuli. 

When the template $\mu$ is unknown, as is typical in practice, for each pair $(i,j)$, the warping functions $\alpha_i$ and $\xi_j$  are individually not identifiable in a nonparametric (infinite-dimensional) specification of the class $\Gamma$, and error $E_i$ without restrictions (e.g., rank-one error model). It is hence not possible to estimate both $\alpha_i$ and $\xi_j$ since their respective variabilities are confounded with the error $E_i$; this implies that it is not possible to decompose the overall phase variation given by $\{\gamma_{ij}\}$ in $\{F_i\}$ into the cross-observation and cross-component phase variations. 

In view of this, our registration procedure will first estimate the cross-component warping functions for each observation $i\in [n]$ and then use them to estimate $\gamma_{ij}$. Evidently then, the \emph{cross-component warping functions $\xi_1,\ldots,\xi_K$ will depend on $i$}. The key challenge arises from the need to iterate between computing the template components $\mu_1, \ldots,\mu_K$ and warping functions $\{\gamma_{ij}\}$ by explicitly incorporating spatial dependence between the, unobserved but observation dependent, cross-component warping functions $\xi_{i1},\ldots,\xi_{1K}$.  In Section \ref{sec:expwarpmod}, we will consider a simplified geometry of $\Gamma$ under a suitable transformation, that will further clarify the dependence of the cross-component phases on the observations. 

\section{Overview of elastic functional data registration}
\label{sec:univfundata}

The proposed algorithm is based on a componentwise spatially-penalized registration of $f_{1j}, \ldots,f_{nj}$ using the metric-based elastic functional data analysis framework \citep{srivastava2011registration,srivastava2016functional} of univariate functions. The metric will dually enable us to propose a novel penalty term that quantifies the spatial correlation between the unobserved cross-component phases $\{\xi_j\}$, unaffected by the cross-observation phase $\{\alpha_i\}$. We begin with a brief review of the elastic framework, and for later comparison with the proposed algorithm,  also review universal registration of curves under the elastic framework \citep{srivastava2016functional} where all components are assumed to have identical temporal variation.  
\subsection{Univariate functions}
\label{sec:univariatefunctions}

We consider the representation space of univariate functional data objects to be $\mathcal F:=\{f:[0,1]\to  \mathbb R\mid f \text{ is absolutely continuous}\}$. The group of warping functions representing phase is $\Gamma=\{\gamma:[0,1]\to[0,1]\mid\gamma(0)=0,\gamma(1)=1,\dot\gamma>0\}$. For any $f\in \mathcal F$, $\gamma\in \Gamma$, the warping of $f$ by $\gamma$ is given by the group action of composition, $f\circ\gamma$. The group-theoretic formulation of phase enables a definition of the amplitude of a function $f$ as the equivalence class $[f]:=\{f\circ \gamma\mid\gamma\in \Gamma\} \subset \mathcal F$, known as its orbit under the action of $\Gamma$; thus, $f\circ \gamma \in [f]$ has the same amplitude as $f$ for each $\gamma \in \Gamma$. The amplitude space then is the quotient space $\mathcal F/\Gamma:=\{[f]\mid f\in\mathcal F\}$.  

Separating amplitude and phase requires a metric on the amplitude space $\mathcal F/\Gamma$. A convenient way to define one is through a metric $d$ on $\mathcal{F}$ that is invariant to simultaneous warping: for every $\gamma\in\Gamma,\  d(f_1,f_2)=d(f_1\circ\gamma,f_2\circ\gamma)$. The standard $\mathbb L^2$ metric fails to be invariant and \citet{srivastava2011registration} thus proposed to use the extended Fisher-Rao (eFR) metric. While direct use of this metric for amplitude-phase separation is difficult in practice, the {square-root slope transform} can be used to `flatten' the complicated eFR metric on $\mathcal F$ to the standard $\mathbb L^2$ metric on the transformed space. The transform maps $f \mapsto Q(f)=q:=\dot f|\dot f|^{-1/2}$ ($\dot f$ is the time derivative of $f$). Given $f(0)$, $Q$ is bijective with inverse $Q^{-1}(q,f(0))(t)=f(t)=f(0)+\int_0^tq(u)|q(u)|du$. Henceforth, for any $f\in{\cal F}$, we will refer to $q=Q(f)$ as its square-root slope function (SRSF). 

The transformed space $Q(\mathcal F)$ is a subset of $\mathbb L^2([0,1],\mathbb R)$, and is denoted by $\mathcal Q$. Under $Q$, the eFR metric on $\mathcal F$ maps to the standard $\mathbb L^2$ metric on $\mathcal Q$, and thus analysis of SRSFs can be carried out using standard Hilbert space machinery. Warping of $f \in \mathcal F$ by $\gamma\in\Gamma$ induces the warping action $q\odot \gamma:=(q\circ\gamma){\dot\gamma}^{1/2}$ on $\mathcal Q$ equipped with the $\mathbb L^2$ metric, and the action is by isometries, i.e., for $q_1,\ q_2\in\mathcal{Q}$ and $\gamma\in\Gamma$, $\|q_1-q_2\|=\|q_1\odot\gamma-q_2\odot\gamma\|$, since $\|q\odot\gamma\|=\|q\|$ for every $\gamma \in \Gamma,\ q \in \mathcal Q$; $\|\cdot\|$ denotes the $\mathbb L^2$ norm. 


\subsubsection{Pairwise registration}
\label{sec:pairwiseunivariate}

Given two functions $f_1,\ f_2\in\mathcal{F}$, amplitude and phase separation through pairwise registration or alignment of $f_2$ to $f_1$, or vice versa, is formulated as the determination of the relative phase of $f_2$ with respect to $f_1$ ($q_1= Q(f_1),\ q_2=Q(f_2)$):
\begin{equation}\label{relphase}
\gamma^*=\underset{\gamma\in\Gamma}{\arg \min} \ \|q_1-q_2\odot\gamma\|^2=\underset{\gamma\in\Gamma}{\arg \min} \ \int_0^1 |q_1(t)-(q_2\odot\gamma)(t)|^2 dt. 
\end{equation}
The minimization problem in \eqref{relphase} is typically solved using the dynamic programming algorithm. To further regularize pairwise registration, one can instead solve the penalized optimization problem given by

\begin{equation}\label{eq:objfn1D}
\gamma^*=\underset{\gamma\in\Gamma}{\arg \min} \ \{\|q_1-q_2\odot\gamma\|^2+\lambda \|\sqrt{\dot\gamma}-1\|^2\},
\end{equation}
where $\lambda$ is the regularization parameter. The penalty is defined as the squared $\mathbb L^2$ distance between the SRSFs of $\gamma$ and the identity warping function $\gamma_{id}$, where $\gamma_{id}(t)=t\ \forall\ t$. It is evident that, depending on the magnitude of $\lambda$, this penalty forces the estimated phase to be close to the identity element, i.e., no warping. 

\subsubsection{Multiple registration}
\label{sec:multreg}

Amplitude-phase separation for a sample of functions $f_i,\ i\in [n],\ n>2$ is carried out with respect to a common template function that must also be estimated. Let $q_i,\ i\in [n]$ denote the corresponding SRSFs. A template for multiple registration is estimated via
\begin{equation}\label{multreg}
    \hat\mu=\underset{\mu\in\mathcal{Q}}{\arg \min}\sum_{i=1}^n\min_{\gamma_i\in\Gamma}\|\mu-q_i\odot\gamma_i\|^2.
\end{equation}
Then, multiple registration is carried out via pairwise registration of each of $q_i,\ i\in [n]$ with respect to $\hat\mu$ using \eqref{relphase}, resulting in the relative phases $\gamma_i^*,\ i\in [n]$.
The minimizer $\hat\mu$ of \eqref{multreg} is not unique, with any element of the orbit $[\hat\mu]$ resulting in the same value of the cost function due to the isometric action of $\Gamma$. Thus, for identifiability, we select $\hat\mu\in [\hat\mu]$ such that the relative phases $\gamma_i^*,\ i\in [n]$ average to $\gamma_{id}$; see \cite{srivastava2011registration} for details on how to compute an average of a sample of warping functions.

\subsection{Componentwise and universal registration}
\label{sec:commonphase}
Let $\tilde{\mathcal F}=\{F:[0,1]\to  R^K\mid F \text{ is absolutely continuous}\}$ denote the the space of multivariate functional data. Each multivariate function $F$ contains $K$ univariate components $f_1,\dots,f_K\in\mathcal{F}$. Independent, componentwise registration of multivariate functional data is to apply the multiple registration given by \eqref{multreg} to functions in each component independently. This fails to account for any correlation between $\gamma_{ij}$ and $\gamma_{ik}$ for every $i \in [n]$ and $j,\ k \in [K] $. 

At the other end of the spectrum is universal registration, which treats each $F_i$ as a parameterised curve $t \mapsto (f_{i1},\ldots,f_{iK})^\top$ in $\mathbb R^K$, and thus assumes the same relative phase for all components. Registration under such a setup is again available through the elastic framework under a suitable transformation. With a slight abuse in notation, let $Q(F)=q=\dot F|\dot F|^{-1/2}$ denote the SRSF of $F$, where $\dot F$ is the componentwise time derivative of $F$ and $|\cdot|$ is the Euclidean norm in $\mathbb R^K$; each SRSF in this case is a function $q:[0,1]\to \mathbb R^K$ and the space of such SRSFs is denoted by $\tilde{\mathcal Q}\subset \mathbb L^2([0,1],\mathbb R^K)$. Then, given $q_1=Q(F_1)$ and $q_2=Q(F_2)$, the relative phase of $F_2$ with respect to $F_1$ is given by
\begin{equation}\label{pairwisemultivariate}
    \gamma^*=\underset{\gamma\in\Gamma}{\arg \min} \ \|q_1-q_2\odot\gamma\|^2=\underset{\gamma\in\Gamma}{\arg \min} \ \int_0^1 |q_1(t)-(q_2\odot\gamma)(t)|^2 dt,
\end{equation}
where again $|\cdot |$ denotes the Euclidean norm in $\mathbb{R}^K$ and $q\odot\gamma$ is applied componentwise. A penalized version of registration can also be implemented in this case by appropriately adapting the optimization problem defined in \eqref{eq:objfn1D}. Further, multiple registration, via estimation of a template $\hat\mu\in\tilde{\mathcal Q}$, and pairwise registration of each function to $\hat\mu$ via \eqref{pairwisemultivariate}, follows the approach defined for univariate functional data in Section \ref{sec:multreg}.

\section{Registration of multivariate functions with spatially dependent cross-component phase}
\label{sec:proposedmethod}

We propose a penalized multiple registration method for multivariate functional data wherein cross-component phase in each observation is spatially correlated. Let $F_i=(f_{i1},\dots,f_{iK})^\top \in\tilde{\mathcal F},\ i\in [n]$ denote a multivariate functional data sample; we assume that $F_i,\ i \in [n]$ are independent. Each component function $f_{ij},\ i \in [n],\ j \in [K]$ is assumed to be an element of $\mathcal{F}$, and the components $f_{ij},\ j \in [K]$, for a fixed $i$,  have spatially dependent phase. Further, let $q_{ij}=Q(f_{ij})\in\mathcal{Q},\ i \in [n],\ j \in [K]$ denote the SRSFs of the component functions in each observation. Registration of multivariate functional data requires estimation of the overall phase (composition of cross-observation and cross-component phase) for each component in each observation, $\gamma_{ij}\in\Gamma,\ i \in [n],\ j \in [K]$. This facilitates simultaneous synchronization of the component functions across $i$ and $j$. 

As a compromise between the two extreme settings of independent componentwise and universal registration described in Section \ref{sec:commonphase}, we propose a spatially penalized registration approach that takes spatial cross-component phase correlation into account, but allows each component to have its own phase. As described in Section \ref{setup}, within each $F_i$ the dependence between the component functions $f_{i1}, \ldots, f_{iK}$ arises through spatial correlation in the cross-component phases $\xi_1,\ldots, \xi_K$. 

\subsection{Spatially penalised componentwise registration}

Our approach is to carry out cross-observation registration through a modification of the multiple registration procedure in \eqref{multreg} using a spatially-informed penalty term. For a fixed function component $j$, the registration of functions $f_{ij},\ i \in [n]$, with SRSFs $q_{ij},\ i \in [n]$, amounts to determination of $\gamma_{ij},\ i \in [n]$. However, for each $i \in [n]$, we note that $\gamma_{ij}$ is spatially correlated with $\gamma_{il},\ l \neq j,\ l \in [K]$, through the spatial correlation between the latent $\xi_j$ and $\xi_l,\ l \neq j$. 

We thus consider a penalised modification of \eqref{multreg} wherein the penalty term for each $i \in [n]$  depends on the phases $\gamma_{i1},\ldots,\gamma_{i(j-1)},\gamma_{i(j+1)},\ldots,\gamma_{iK}$. Our choice is to define the penalty term using (an estimate of) the conditional mean of $\gamma_{ij}$ given $\{\gamma_{il}\}_ {l \neq j}$: we wish to discourage $\gamma_{ij}$ from assuming values that are far away from its spatially weighted conditional mean. We accordingly consider the following optimization problem for each component $j \in [K]$:
\begin{equation} \label{eq:objfn_pen}
(\gamma_{1j}^*,\dots,\gamma_{nj}^*,\hat\mu_j)=\underset{\gamma_{1j},\dots,\gamma_{nj}\in\Gamma,\ \mu_{j}\in\mathcal{Q}}{\arg \min}\ \sum\limits_{i=1}^n \left\{\|\mu_j-q_{ij}\odot\gamma_{ij}\|^2+\lambda \thinspace d^2(\gamma_{ij},\tilde \gamma_{ij})\right\},
\end{equation}
where $\mu_j$ is the template for registration in component $j$, $\lambda>0$ is a penalty parameter,  and $d$ is a distance on $\Gamma$; the warping function $\tilde \gamma_{ij}$ is an estimate of the conditional mean of $\gamma_{ij}$ given $\{\gamma_{il}\}_{ l \neq j}$, which we will define next in Section \ref{sec:expwarpmod}. 
 
 The first term in the objective function in \eqref{eq:objfn} provides a measure of synchronization for component $j$, across observations $i$, with respect to the template $\mu_j$. The second term, a penalty on phase, measures the distance between the estimated phase $\gamma_{ij}$ and a target $\tilde \gamma_{ij}$ determined by the correlated phase in the other components. The regularization penalty attempts to preserve the phase-induced correlation structure in the aligned components. 
If the function components are assumed to be independent, and $\tilde \gamma_{ij}=\gamma_{id}$, the identity warping function, for all $i$ and $j$, the proposed approach is equivalent to $K$ independent univariate penalized registration problems, as specified in \eqref{eq:objfn1D}, with $\mu_j$ acting as the template for each component. 

\subsection{Penalty using spatial model for cross-component phase}
\label{sec:expwarpmod}

We first discuss the choice of distance $d$ on $\Gamma$ in the penalty term in \eqref{eq:objfn_pen}. Elements of the function space $\Gamma$ can be viewed as differentiable, increasing distribution functions of random variables on $[0,1]$, and thus constitute a nonlinear, convex set. A simplified geometric structure compatible with the $\mathbb L^2$ norm preserving action under the operation $\odot$ is available under the SRSF transform, under which $\gamma \to Q(\gamma)=\dot \gamma ^{1/2}=:\psi$. The SRSF map $Q$ is bijective, 
and each $\psi$ in its image corresponds to a square-root probability density since $\int_0^1 \psi^2(t) \text{d}t=1$. As a consequence, the set $\Psi:=\{Q(\gamma)=:\psi:[0,1]\to \mathbb R_+\mid\gamma\in \Gamma\}$ can be identified with the positive orthant of the unit sphere in $\mathbb L^2([0,1],\mathbb R)$. While the natural candidate for distance $d$ on $\Gamma$ is the intrinsic arc-length distance on $\Psi$, we use the extrinsic distance
\begin{equation*}
d(\gamma_1,\gamma_2):=\|Q(\gamma_1)-Q(\gamma_2)\|=\|\psi_1-\psi_2\|, 
\end{equation*}
where $\|\cdot\|$ is the standard $\mathbb L^2$ norm. Our choice is linked to the choice of $\tilde \gamma_{ij}$ in the penalty term. Observe that with $d$ defined as above on $\Gamma$, we have that, for every $\alpha,\ \gamma_1,\ \gamma_2 \in \Gamma$, 
\begin{equation}
	\label{eq:isometry}
d(\gamma_1 \circ \alpha,\gamma_2 \circ \alpha)=\|Q(\gamma_1 \circ \alpha)-Q(\gamma_2 \circ \alpha)\|=\|\psi_1 \odot \alpha -\psi_2 \odot \alpha \|=\|\psi_1-\psi_2\|,
\end{equation}
and $\Gamma$ acts on itself by isometries under the SRSF map. We thus combine the cross-observation phase $\alpha_i$ and cross-component phase $\xi_j$ using their SRSFs to obtain the SRSF $\psi_{ij}$ of $\gamma_{ij}$ as
\begin{equation}
\label{eq:cross}
\psi_{ij}:=Q(\xi_j \circ \alpha_{i})
=Q(\xi_j) \odot \alpha_i.
\end{equation}
As described in Section \ref{setup}, our registration procedure will estimate $\psi_{ij}$ by first estimating the spatially correlated cross-component phases for \emph{each} observation $i\in [n]$ since $\alpha_i$ and $\xi_{j}$ are individually not estimable; in other words, we wish to induce dependence in the generative model \eqref{eq:model} between the confounded cross-component and cross-observation phases. We achieve this using the SRSF transform of $\gamma_{ij}$ in the manner defined in \eqref{eq:cross}. 

Under the SRSF transform of $\Gamma$, the population conditional expectation $E[\psi_{ij}\mid \psi_1,\ldots,\psi_{i(j-1)},\psi_{i(j+1)},\ldots,\psi_{iK}]$ is an ideal choice for $\psi_{ij}$. As with spatially correlated real- or vector-valued observations in traditional statistics modeled using random fields, a viable estimate of the conditional expectation in this context is the spatial interpolant or \emph{kriging estimate}  at location $\mathbf s_j$ of a functional random field 
$$
\big\{\psi_{\mathbf s}:=Q(\xi_{\mathbf s}) \odot \alpha,\ \mathbf s \in \mathcal D\big \}
$$
assuming values in $\Psi$, conditioned on its values at locations $\mathbf s_1,\ldots, \mathbf s_{j-1}, \mathbf s_{j+1},\ldots \mathbf s_{K}$. In other words, the random field $\{\psi_{\mathbf s}\}$ is derived by time-warping the values assumed by another functional random field $\{\xi_{\mathbf s},\ \mathbf s \in \mathcal D\}$ with a fixed $\alpha \in \Gamma$. 

Computing the kriging estimate of $\{\psi_{\mathbf s}\}$ at $\mathbf s_j$ requires estimation of $Q(\xi_{i1}),\ldots,Q(\xi_{iK})$. Suppose that the components $f_{ij},\ i \in[n],\ j \in [K]$ are indexed by spatial locations $\mathbf s_j\in\mathcal D,\ j \in [K]$. Denote by $\psi_{ij}$ the SRSFs of the warping functions $\gamma_{ij}$, which, for every $i$, are viewed as values assumed by the random field $\{\psi_{\mathbf s}:\mathbf s\in\mathcal D\}$ defined above at spatial locations $\mathbf s_1,\ldots, \mathbf s_K$.  Then, under the assumption that the functional random field $\{\psi_{\mathbf s}\}$ is second-order stationary and isotropic, we can consider the \emph{phase trace-variogram}, defined by \cite{guo2020variograms} as
\[
V(h)=\frac{1}{2}\int_0^1 E\left[\psi_{\mathbf s}(t)-\psi_{\mathbf s'}(t))\right]^2 \text{d}t=\frac{1}{2}E(\|\psi_{\mathbf s}-\psi_{\mathbf s'}\|^2),
\]
using Fubini's theorem, where $h=|\mathbf s- \mathbf s'|$ and $|\cdot|$ is the Euclidean norm on the spatial domain $\mathcal{D} \subset \mathbb R^p$. For a fixed observation $i$, the estimator of $V(h)$ is given by
\begin{align}\label{eq:emp_vargm}
\hat V_{i}(h)=\frac{1}{2|N(h)|}\sum \limits_{a,b\in N(h)} \|\hat \psi_{ia}-\hat \psi_{ib}\|^2,
\end{align}
where $N(h)=\{( \mathbf  s_a, \mathbf  s_b)\mid a,b =1,\ldots,K,\ | \mathbf  s_a-  \mathbf s_b|=h \}$. For irregularly spaced data, $N(h)$ can be modified to $N_{\epsilon}(h) = \{( \mathbf  s_a, \mathbf  s_b):| \mathbf  s_a- \mathbf  s_b|\in (h-\epsilon,h+\epsilon)\}$ for a small $\epsilon >0$.

Despite the fact that the spatially correlated cross-component phases $\xi_1,\ldots,\xi_K$ are unobserved and confounded with the cross-observation phases $\alpha_1,\ldots,\alpha_n$, we can access their correlation structure by estimating $\gamma_{ij}$ using $V(h)$ and $\hat V(h)$. To see this, note that due to the isometry property in \eqref{eq:isometry}, 
\begin{align*}
E(\|\psi_{\mathbf s}-\psi_{\mathbf s'}\|^2)=E(\|Q(\xi_{\mathbf s})\odot \alpha-Q(\xi_{\mathbf s'})\odot \alpha)\|^2)
=E(\|Q(\xi_{\mathbf s})-Q(\xi_{\mathbf s'})\|^2),
\end{align*}
and  $V$ thus equals the phase-trace variogram of the functional random field $\{Q(\xi_{\mathbf s})\}$; a similar argument applies to the estimate $\hat V$ as well. We thus note that the phase-trace variogram is invariant to (simultaneous) warping  \citep[Lemma 1]{guo2020variograms}, and this crucial observation motivates our use of the extrinsic distance on $\Psi$ and the functional random field $\{\psi_{\mathbf s}\}$ with values in $\Psi$. 

We now move on to computing the kriging estimate $\tilde \psi_{ij}$ using the variogram estimate $\hat V$. For component $j$ in observation $i$, given phase for the other components $\{\psi_{il}\}_{l\neq j}$, the conditional mean phase is given by the weighted average
 \begin{equation}\label{phase_pred}
\tilde \psi_{ij}=\sum\limits_{l\neq j}\zeta_{ijl} \psi_{il},\quad \text{where } \sum\limits_{l\neq j}\zeta_{ijl}=1,\ \zeta_{ijl}\geq 0.
\end{equation}
The coefficient vector $\bm \zeta_{ij} = \{\zeta_{ijl}\}_{l\neq j}$ is implicitly defined as the minimizer of 
 \begin{align}\label{eq:phaseerror}
 \bm \zeta_{ij} \mapsto E\|\tilde \psi_{ij}-\psi_{ij}\|^2.
 \end{align}
\citet{guo2020variograms} show that $\bm \zeta_{ij}$ can be estimated using a quadratic optimization problem that depends on the trace-variogram $V(h)$ and pairwise spatial distances $|\mathbf s_a-\mathbf s_b|$, for $a,b=1,\ldots,K$. The estimator $\tilde \psi_{ij}$ is subsequently normalized using $\tilde \psi_{ij}\to\tilde \psi_{ij}/\|\tilde \psi_{ij}\|$ to correspond to a valid warping function.

Summarily, for the $i$th observation, if $\tilde \psi_{ij}$ denotes the kriging or predicted value at $\mathbf s_j$, the penalty term in \eqref{eq:objfn_pen} becomes $d(\gamma_{ij},\tilde \gamma_{ij})=\|\psi_{ij}-\tilde \psi_{ij}\|$,
and the optimization problem for registration in \eqref{eq:objfn_pen} with the spatial penalty assumes the specific form
 \begin{equation} \label{eq:objfn}
 	(\gamma_{1j}^*,\dots,\gamma_{nj}^*,\hat\mu_j)=\underset{\gamma_{1j},\dots,\gamma_{nj}\in\Gamma,\ \mu_{j}\in\mathcal{Q}}{\arg \min}\ \sum\limits_{i=1}^n \left\{\|\mu_j-q_{ij}\odot\gamma_{ij}\|^2+\lambda\|\psi_{ij}-\tilde\psi_{ij}\|^2\right\}.
 \end{equation}
 \subsection{Registration algorithm and implementation details}

\begin{figure}[!t]
	\centering
	\includegraphics[scale=0.5]{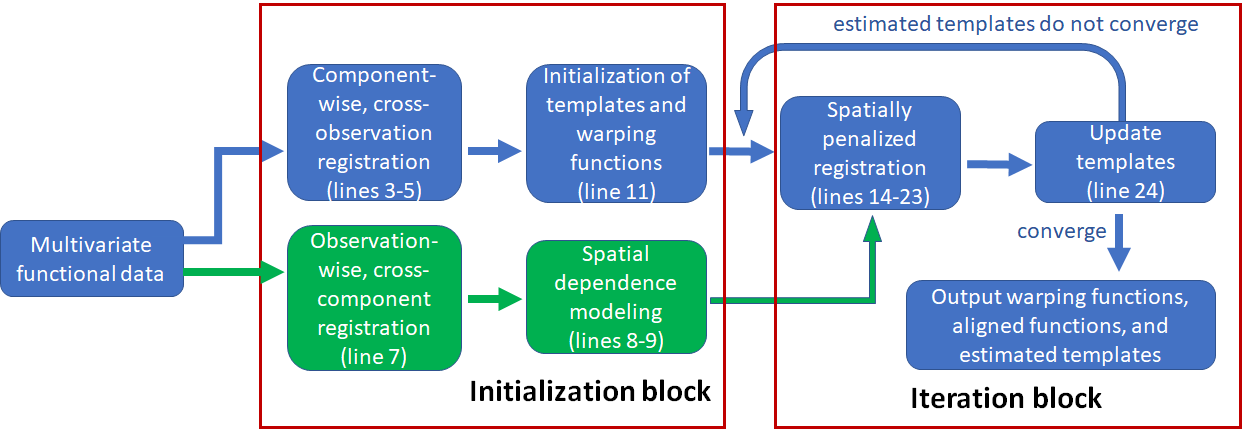}
	\caption{\small High-level overview of Algorithm 1.}
	\label{flowchart}
\end{figure}

The optimization problem for registration in \eqref{eq:objfn} is solved in an iterative fashion in Algorithm \ref{alg:jointregist} given below, where tools from Section \ref{sec:expwarpmod} are used to model the spatially dependent cross-component phase within each observation. 
Figure \ref{flowchart} provides a high-level diagrammatic representation of Algorithm \ref{alg:jointregist}. 
More specifically, given component functions $\{f_{ij},\ i \in [n],\ j \in [K]\}$ of multivariate functional data $F_1,\ldots,F_n$, we first transform them to obtain their SRSFs $\{q_{ij}\}$. Then, for a fixed penalty parameter $\lambda>0$, Algorithm \ref{alg:jointregist} can be decomposed into two blocks: 
\newpage
\begin{description}[leftmargin=*]
	\item \textbf{Initalization}:
	\begin{enumerate}[label=(\roman{*}), leftmargin=*]
		\itemsep 0em
	\item Obtain initial template component functions $\hat \mu^{(0)}_1, \ldots, \hat \mu^{(0)}_K$ by performing independent componentwise registration of $\{q_{1j}, \ldots, q_{nj}\}$ for each $j \in [K]$ given by \eqref{multreg};
	\item initialize the phase functions to the identity warping by initializing their SRSFs to $\psi^{(0)}_{ij} \equiv 1,\ i \in [n],\ j \in [K]$;
	\item for each observation $i \in [n]$, implement multiple registration of $\{q_{i1}, \ldots, q_{iK}\}$ using \eqref{multreg} to estimate cross-component phase $\xi_{i1},\ldots,\xi_{iK}$;
	\item for each observation $i \in [n]$, compute weights $\zeta_{i1}, \ldots,\zeta_{iK}$ required for the kriging estimate using the estimated phase trace variogram $\hat{V}_i$.
	\end{enumerate}
	\item \textbf{Iteration}: Starting with the initial values, obtain phases $\{\hat \gamma_{ij}\}$ and template components $\{\hat \mu_j\}$ by iterating and updating, until convergence.
	\begin{enumerate}[label=(\roman{*}), leftmargin=*]
				\itemsep 0em
		\item Compute $\tilde \psi_{ij}$ in penalty using kriging coefficients 
		\{$\zeta_{ij\ell}\}_{\ell\neq j}$ and phases $\{\hat \gamma_{i\ell}\}_{\ell \neq j} $;
		\item solve $\hat\gamma_{ij}=\underset{\gamma\in\Gamma}{\arg \min}\ \left\{\|\hat\mu_j-q_{ij}\odot\gamma\|^2+\lambda\|\sqrt{\dot\gamma}-\tilde\psi_{ij}\|^2\right\}$;
 \item compute template components $\{\hat \mu_j\}$ by averaging $\{q_{ij} \odot \hat \gamma_{ij}\}$ over $i$. 
\end{enumerate}
\end{description}

\begin{algorithm}[!t]
\caption{Spatially penalized registration of multivariate functions}
\label{alg:jointregist}
\begin{algorithmic}[1]
\State \textbf{Input}: SRSFs $q_{ij},\ i \in [n],\ j \in [K]$ and regularization parameter $\lambda$.
\State \textbf{Output}: Estimated componentwise template functions $\hat\mu_j,\ j \in [K]$, and warping functions $\hat \gamma_{ij},\ i \in [n],\ j \in [K]$.
\For{ $j = 1$ to $K$}
\State Multiple alignment of $q_{ij},\ i \in [n]$ via \eqref{multreg} to initialize the template $\hat\mu_{j}^{(0)}$.
\EndFor
\For{ $i = 1$ to $n$}
\State Multiple alignment of $\{q_{ij},\ j \in [K]\}$ via \eqref{multreg} to estimate transformed cross-component phase $\{\hat\xi_{ij},\ j \in [K]\}$;
\State Estimation of trace-variogram $\hat V_i(h)$ in \eqref{eq:emp_vargm} using $\{\hat\xi_{ij},\ j \in [K]\}$;
\State Estimation of coefficient vector $\bm \zeta_{ij}$ using $\hat V_i$ (Section \ref{sec:expwarpmod}).
\EndFor
\State Set $z=0$, $\epsilon_1>0$ and $\hat\psi_{ij}^{(0)}(t)=1,\ i \in [n],\ j \in [K]$ with $\hat\mu_j^{(0)} = n^{-1}\sum_{i=1}^n q_{ij}$, $j \in [K]$. 
\While{$z=0$ or $\sum_{j=1}^K \|\hat\mu_{j}^{(z)}-\hat\mu_{j}^{(z-1)}\| >\epsilon_1$}
\For{ $i = 1$ to $n$}
\State Initialization of $k=0$, $\epsilon_2>0$ small;
\While{$k=0$ or $\sum_{j=1}^K \|\hat\psi_{ij}^{(k)}-\hat\psi_{ij}^{(k-1)}\|^2  >\epsilon_2$}
\For{ $j = 1$ to $K$}
\State  Estimation of $\tilde \psi_{ij}$ using $\bm \zeta_{ij}$ and $[\hat\psi^{(k+1)}_{i1},...,\hat\psi^{(k+1)}_{i(j-1)},\hat\psi^{(k)}_{i(j+1)},...,\hat\psi^{(k)}_{iK}]$;
\State Solve $\hat\gamma_{ij}=\underset{\gamma\in\Gamma}{\arg \min}\ \left\{\|\hat\mu^{(z)}_j-q_{ij}\odot\gamma\|^2+\lambda\|\sqrt{\dot\gamma}-\tilde\psi_{ij}\|^2\right\}$ and compute SRSF $\hat \psi^{(k+1)}_{ij}$ of $\hat\gamma_{ij}$.
\State Set $k = k+1$.
\EndFor
\EndWhile
\State Set $\hat \gamma_{ij} = \hat \gamma_{ij}^{(k)}$
\EndFor
\State Set $\hat\mu_{j}^{(z+1)}=\frac{1}{n}\sum_{i=1}^n (q_{ij}\odot \hat \gamma_{ij}),\ j \in [K]$;
\State Set $z=z+1$. 
\EndWhile
\end{algorithmic}
\end{algorithm}

Algorithm \ref{alg:jointregist} contains nested while loops (lines 12-26). The inner loop indexed by $k$ (lines 15-21) updates the estimated warping functions given the templates for all components. The outer loop index by $z$ (lines 12-26) updates the template for each component given the estimated warping functions. The convergence of Algorithm \ref{alg:jointregist} is assessed empirically using simulation studies in Section \ref{sec:simu}.

\section{Simulation studies} \label{sec:simu}

We conduct simulation studies to assess the performance of the proposed multiple registration approach for multivariate functional data. Specifically, we estimate the componentwise template functions $\{\mu_j,\ j \in [K]\}$ using three different approaches: (1) the proposed method, which utilizes spatial cross-component phase correlation in penalized registration (Section \ref{sec:proposedmethod} and Algorithm \ref{alg:jointregist}), (2) independent componentwise registration (Section \ref{sec:univariatefunctions} and Section 3.4 in \citep{srivastava2011registration}), and (3) universal registration wherein each component is assumed to have the same phase (Section \ref{sec:commonphase}).


\subsection{Data generating model}
\label{sec:datagen}

We consider multivariate functional data wherein dependence between function components is induced by spatial correlation. In this setting, each component of a simulated observation is indexed by a spatial coordinate $\mathbf s\in\mathbb R^p$; component $j$ for observation $i$ is denoted by $f_{i,\mathbf s_j}$. Component functions of the data are generated using the model:
$$f_{i,\mathbf s_j}(t)=(\mu_{\mathbf s_j}+e_{i,\mathbf s_j})\circ(\xi_{i,\mathbf s_j}\circ \alpha_i)(t),
\quad t\in[0,1],\  i\in[n],\ j\in [K].$$
\noindent In the model, $\mu_{\mathbf s_j}\in\mathcal{F}$ denotes the template for component $j$ indexed by spatial location $\mathbf s_j$, which is the object that we wish to estimate through a registration procedure. The random functional error $e_{i,\mathbf s_j}$ and cross-component warping functions $\xi_{i,\mathbf s_j}\in\Gamma$ are also indexed by the spatial location. The warping function $\alpha_i\in\Gamma$ denotes the observation specific warping function (common across all components). 

We consider two simulation settings characterised by choice of the component template functions (see below). For both settings, the random phase components are composed of two warping functions: $\alpha_i\in\Gamma$, the phase that is common across components for observation $i$, and $\xi_{i,\mathbf s_j}\in\Gamma$, the cross-component phase within observation $i$. Such a nested structure is similar to the structure in a mixed-effects model. The warping functions $\alpha_i$ are  taken to be cumulative distribution functions (CDFs) of a beta distribution, $\text{Beta}(1,\exp(z_i))$, with random parameter $z_i\sim\text{Unif}[-Z,Z]$. The cross-component phase $\xi_{i,\mathbf s_j}$ is also the CDF of a beta distribution, $\text{Beta}(1,\exp(b_{i,\mathbf s_j}))$,
where $(b_{i,\mathbf s_1},...,b_{i,\mathbf s_K})^\top$ follows the correlated uniform distribution on $[-B,B]$, independently for every $i$; a sample from the correlated uniform distribution can be generated by transforming a correlated multivariate normal sample with mean $(0,\dots,0)^T$ and Matern covariance $C_{Mat}(\cdot,\cdot;1,0.5,\ell)$. The parameters $Z$ and $B$ control magnitudes of the cross-observation and cross-component phase variations; the range parameter is set to $\ell = 0.5\times d_{max}$, where $d_{max}$ is the maximum spatial distance between the simulated sites ${\mathbf s_j}$.

\noindent \underline{\textbf{Simulation setting 1}}: In this setting the component template functions $\{\mu_{\mathbf s_j}\}$ have a specific bimodal form. We consider $K=20$ components where the spatial coordinates $\mathbf s_j$ are generated using a uniform distribution on $[-2,2]^2\subset \mathbb R^2$. The componentwise template functions are generated as $\mu_{\mathbf s_j}(t) = a_{1\mathbf s_j}\exp(-100(t-1/3)^2)+a_{2\mathbf s_j}\exp(-100(t-2/3)^2)$, where $(a_{m,\mathbf s_1},\ldots,a_{m\mathbf s_K})^\top,\ m=1,2$ are independently sampled from a multivariate normal distribution with mean vector $(3,\ldots,3)^\top$ and Matern covariance $C_{Mat}(\cdot,\cdot; \sigma_a^2,0.5,\ell)$; here, $\sigma_a^2$ is the scale parameter, $\ell$ is the range, and the smoothing parameter is fixed to $0.5$. 
The random errors $(e_{i,\mathbf s_1}(t),\ldots,e_{i,\mathbf s_K}(t))^\top$ are generated independently for each observation $i$ in a pointwise manner: for each value of $t$ we generate a sample from the multivariate normal distribution with mean vector $(0,\ldots,0)^\top$ and covariance $C_{Mat}(\cdot,\cdot; \sigma_e^2,0.5,\ell)$. The other covariance parameters are set to $\sigma_a = 1$, $\sigma_e=0.5$ and $\ell = 0.5\times d_{max}$. 

\noindent \underline{\textbf{Simulation setting 2}}:
In this setting, in order to replicate the structure of EEG data, the component template functions $\{\mu_{\mathbf s_j}\}$ are chosen to possess the (typical) shape of EEG signals, i; thus each component is treated as a signal from an EEG electrode. We consider 16 electrode locations on the scalp with three-dimensional coordinates. To imitate the shape variation in EEG signals from different brain regions, we use a more flexible data generating model. The componentwise template functions are generated using 10 B-spline basis functions, $B_m,\ m\in[10]$, of order 4 on the interval $[0,1]$: $\mu_{\mathbf s_j}(t)=\sum_{k=1}^{10}\beta_{k,\mathbf s_j}B_k(t)$. The B-spline basis coefficients $(\beta_{m,\mathbf s_1},\ldots,\beta_{m,\mathbf s_K})^\top,\ m\in[10]$ are generated independently from a multivariate normal distribution with mean vector $(0,\ldots,0)^\top$ and Matern covariance $C_{Mat}(\cdot,\cdot;\sigma_a^2,0.5,\ell)$. Random functional errors, $e_{i,\mathbf s_j}$, are generated in the same way as in Simulation setting 1, but under a lower signal-to-noise ratio: the covariance parameters are set to $\sigma_a = 2$, $\sigma_e=0.5$ or $1$ and $\ell = 0.5\times d_{max}$.

\subsection{Assessing performance of template estimation}

Denote by $\tilde f_{i,\mathbf s_j}$ and $\tilde q_{i,\mathbf s_j}$ the aligned multivariate functions and their SRSFs obtained after registration. We define two performance metrics that quantify the accuracy of template estimation based on the registered functions: 
\begin{equation*}
\text{MSE} = \frac1{Kn}\sum\limits_{j=1}^K\sum\limits_{i=1}^n\|\tilde f_{i,\mathbf s_j}-\mu_{\mathbf s_j}\|^2,
\quad
\text{QMSE} = \frac1{Kn}\sum\limits_{j=1}^K\sum\limits_{i=1}^n\|\tilde q_{i,\mathbf s_j}-Q(\mu_{\mathbf s_j})\|^2,
\end{equation*}
\noindent where $\mu_{\mathbf s_j}$ is the true template for component $j$.  MSE, based on the $\mathbb L^2$ norm on $\mathcal{F}$, is sensitive to phase errors, while QMSE, based on the eFR metric on $\mathcal{F}$, is sensitive to shape differences \citep{srivastava2016functional}. 

We use $n=20,\ 20$ and $K=20,\ 16$, respectively, for the first and second simulation settings described in Section \ref{sec:datagen}; we replicate each simulation 50 times. When implementing the proposed method, we use 4-fold cross-validation to select a value for the regularization parameter $\lambda$. Denote by $\hat\mu^{[-k]}_{\mathbf s_j}$ the estimated template function for component $j$ based on data in all folds except the $k$th one; here, we use $[k]$ to denote the set of observation indices in fold $k$. Then, the value for the regularization parameter $\lambda$ is chosen by minimizing the criterion
$$\frac 1{4Kn}  \sum\limits_{k=1}^4\sum\limits_{j=1}^K\sum\limits_{i\in[k]}\| f_{i,\mathbf s_j}\circ \hat \gamma_{i,\mathbf s_j}-\hat\mu^{[-k]}_{\mathbf s_j}(\lambda) \|^2,$$
where $\hat \gamma_{i,\mathbf s_j} = \arg \min_{\gamma\in \Gamma} \|Q(\hat\mu^{[-k]}_{\mathbf s_j}(\lambda))- Q(f_{i,\mathbf s_j})\odot \gamma\|^2$. That is, we align the functions in the validation set (fold $k$) to the template function estimated using the training set (all folds except $k$). The selected value for the regularization parameter is then used for registration of all observations.

\begin{figure}[!t]
    \centering
    \includegraphics[scale=0.12]{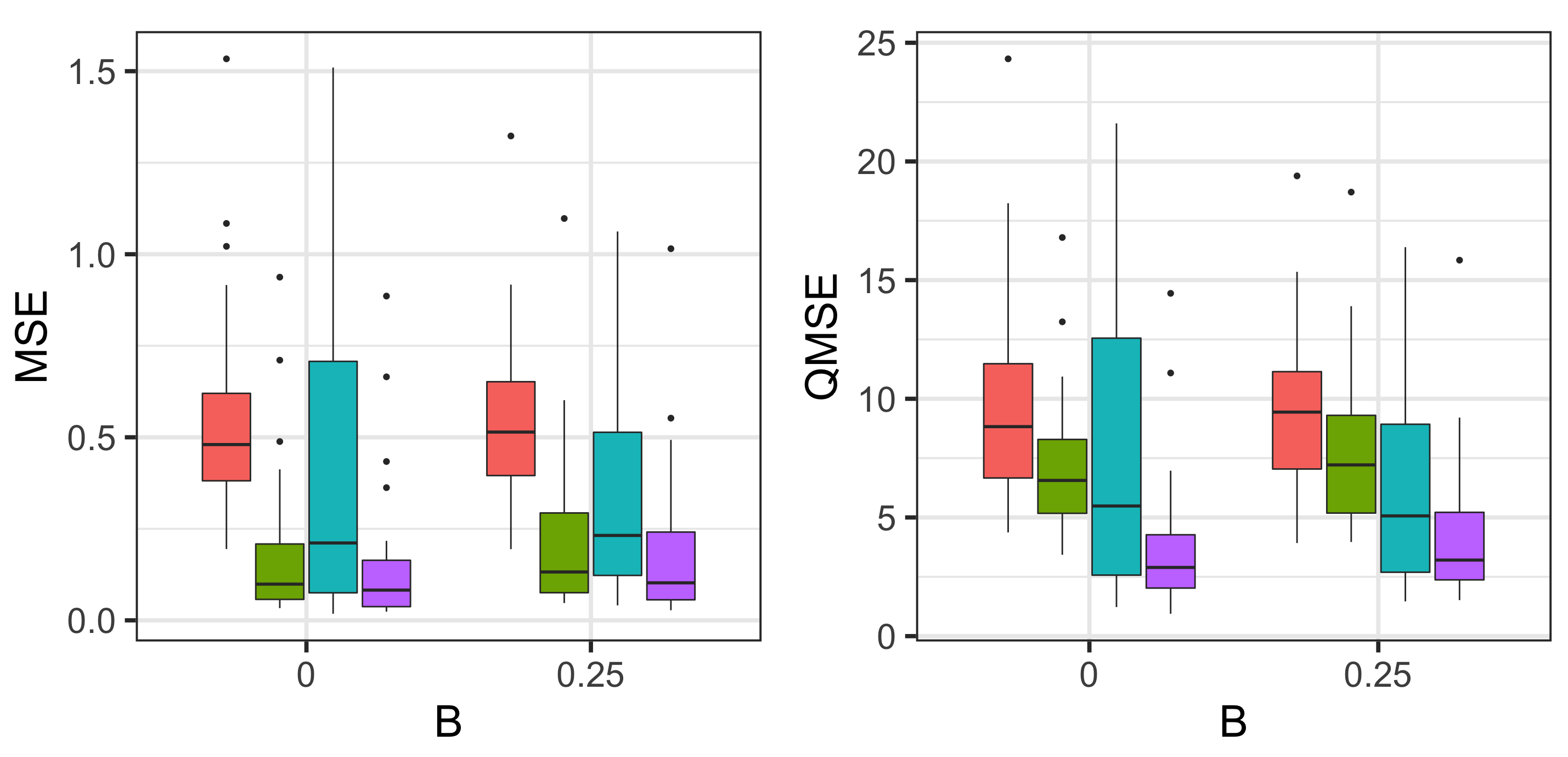}
    \caption{\small Simulation setting 1 with $Z=0.5$ and $\sigma_e = 0.5$. Boxplots of the MSE (left) and QMSE (right) for template functions estimated with no registration (red), componentwise registration (green), universal registration (cyan) and proposed penalized registration (purple). The scale parameter of cross-component phase variation is set to $B=0$ or $B=0.25$.}
    \label{fig:simu_mfd}
\end{figure}

\begin{figure}[!t]
    \centering
    \includegraphics[scale=0.12]{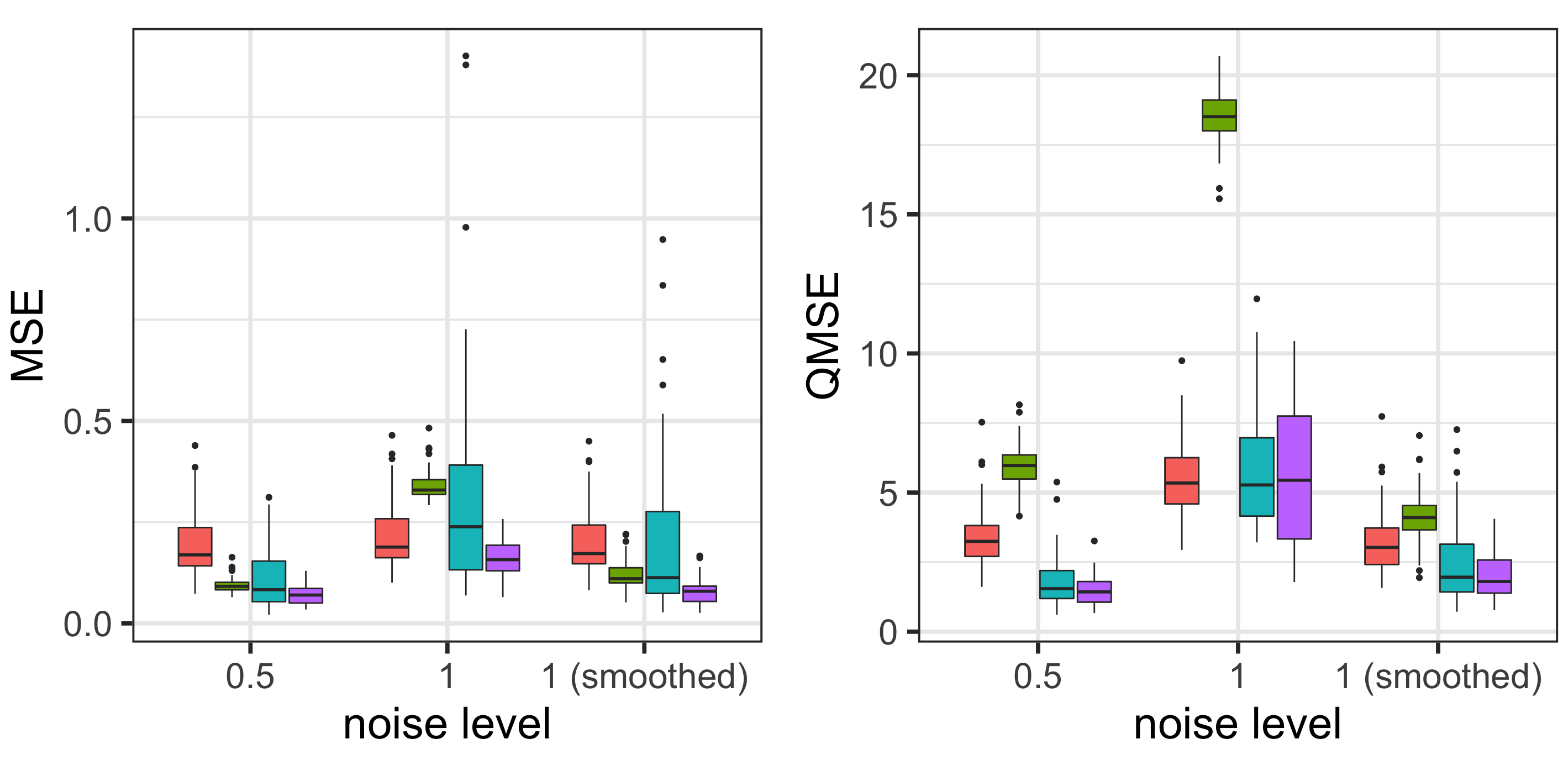}
    \caption{\small Simulation setting 2 with $Z=0.5$ and $B=0$. Boxplots of the MSE (left) and QMSE (right) for template functions estimated with no registration (red), componentwise registration (green), universal registration (cyan) and proposed penalized registration (purple). The noise levels are set to $\sigma_e = 0.5$ or $\sigma_e = 1$. In the case of $\sigma_e = 1$, we also present results after applying smoothing splines with a low value of the smoothing parameter.}
    \label{fig:simu_eeg}
\end{figure}

Results for each simulation setting described in Section \ref{sec:datagen} are summarized in the boxplots shown in Figures \ref{fig:simu_mfd} and \ref{fig:simu_eeg}, respectively. In all cases, we also report values of the error metrics when no registration is applied to the data; this serves as the baseline. 

\noindent \underline{\textbf{Simulation setting 1}}: We observe in Figure \ref{fig:simu_mfd} that both componentwise registration and the proposed method clearly have smaller MSEs than universal registration; further, the mean MSE for the proposed method (0.083) is slightly lower than the mean MSE for the componentwise method (0.099). With respect to QMSE, the performance of componentwise registration is clearly worse than the proposed approach. Interestingly, universal registration outperforms componentwise registration in terms of median QMSE. 

The improvement in registration quality with the proposed method over the componentwise approach is due to the addition of the regularization term. Componentwise registration tends to `overalign' features of the component functions, e.g., peaks and valleys, that are due to the random error, i.e., the estimate of cross-component phase is too flexible. On the other hand, the universal approach performs poorly since all components within an observation are constrained to be identical, i.e., the estimate of cross-component phase is too restrictive. The compromise achieved by the proposed method between the componentwise and universal methods by introducing a regularization penalty that synthesizes spatial phase information from all components within an observation results in improved registration. Finally, we note that the performance of the proposed method is consistently better than the other two approaches, even if no cross-component phase variation exists in the data (the case when $B=0$).

\noindent \underline{\textbf{Simulation setting 2}}: Recall that the setting here is designed to replicate the structure in EEG data. The low signal-to-noise ratio presents significant challenges for all of the registration procedures. Figure \ref{fig:simu_eeg} shows that the componentwise approach, expectedly, is the most sensitive to the random function errors due to its lack of regularization. Despite the good performance of this method in the first simulation setting, the low signal-to-noise ratio when $\sigma_e = 1$ significantly deteriorates the template estimates, especially with respect to QMSE. 
In fact, when $\sigma_e = 1$, the advantages of registration are not obvious since all of the procedures are essentially driven by random noise; regularization with the spatial penalty provides some help, but there is insufficient structure in $\{f_{i,\mathbf s_j}\}$ to inform estimation of the conditional mean phase based on cross-component spatial correlation. In such situations, we recommend slight smoothing of the observed functions prior to registration using any of-the-shelf smoothing procedure (e.g., splines). To examine value in the pre-smoothing, we report additional results in Figure \ref{fig:simu_eeg} after applying smoothing splines, with a small value for the smoothing parameter, to the simulated data when $\sigma_e = 1$. 
The presented results show that the proposed registration method produces more accurate and stable template estimates than the other two registration approaches. 

\begin{figure}[!t]
    \centering
    \includegraphics[width =1\textwidth]{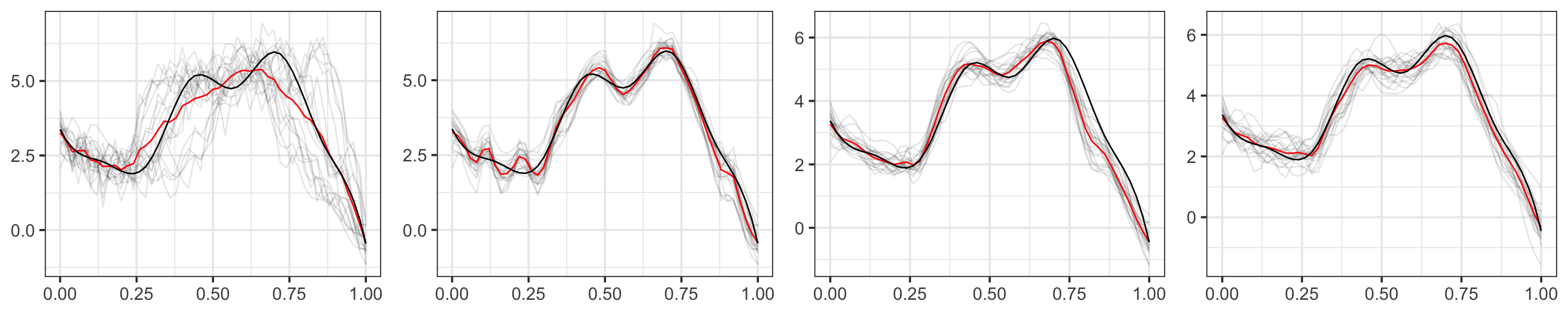}
    \vspace{-.75cm}
    \caption{\small Registration result for a single component in one simulation replicate under Simulation setting 2 with $\sigma_e = 0.5$. Panel 1: Simulated data in grey with cross-sectional average without registration in red. Panels 2-4: Aligned functions in grey with estimated template in red, generated using componentwise registration, universal registration, and the proposed method, respectively. The ground truth template function is shown in black.}
    \label{fig:simu_eeg_example}
\end{figure}

In Figure \ref{fig:simu_eeg_example} we provide results of template estimation for a single component in one simulation run when $\sigma_e=0.5$. The simulated data for this component is shown in grey with the ground truth template in black. The estimated templates are shown in red. In Panel 1 of Figure \ref{fig:simu_eeg_example}, it is evident that when no registration is performed, the resulting template underestimates relevant shape features, e.g., the two peaks and one valley between $t=0.4$ and $t=0.75$; in Panel 2, componentwise registration results in a template that generates additional shape features due to `overalignment' of noise, e.g., the two small peaks and three valleys between $t=0$ and $t=0.3$; in Panel 3, universal registration results in a template that has the `correct' shape, but that is out of phase with respect to the ground truth template, e.g., there is significant lag between $t=0.7$ and $t=1$; Finally, in Panel 4, we note that the proposed approach results in a template that is very similar in shape to the ground truth and is largely in phase. 

\subsection{Assessing convergence of Algorithm \ref{alg:jointregist}}

We empirically assess convergence properties of Algorithm \ref{alg:jointregist} by monitoring the average cost function and the relative change in estimated phase components with respect to the iteration counter. We use a single simulation example generated under Simulation setting 1 with $B=0$. To capture the complete trend of convergence, we remove the stopping rule in the algorithm and complete a fixed number of iterations. Recall that the algorithm has nested while loops. The inner loop updates the template for each component, while the outer loop updates the estimated warping functions given the templates. The number of iterations for the outer loop is set to 20; the number of iterations for the inner loop is set to 10. In our evaluation, the overall number of iterations is defined as the cumulative count of the iterations completed in the inner loop. 

\begin{figure}[!t]
    \centering
    \includegraphics[width =0.8\textwidth]{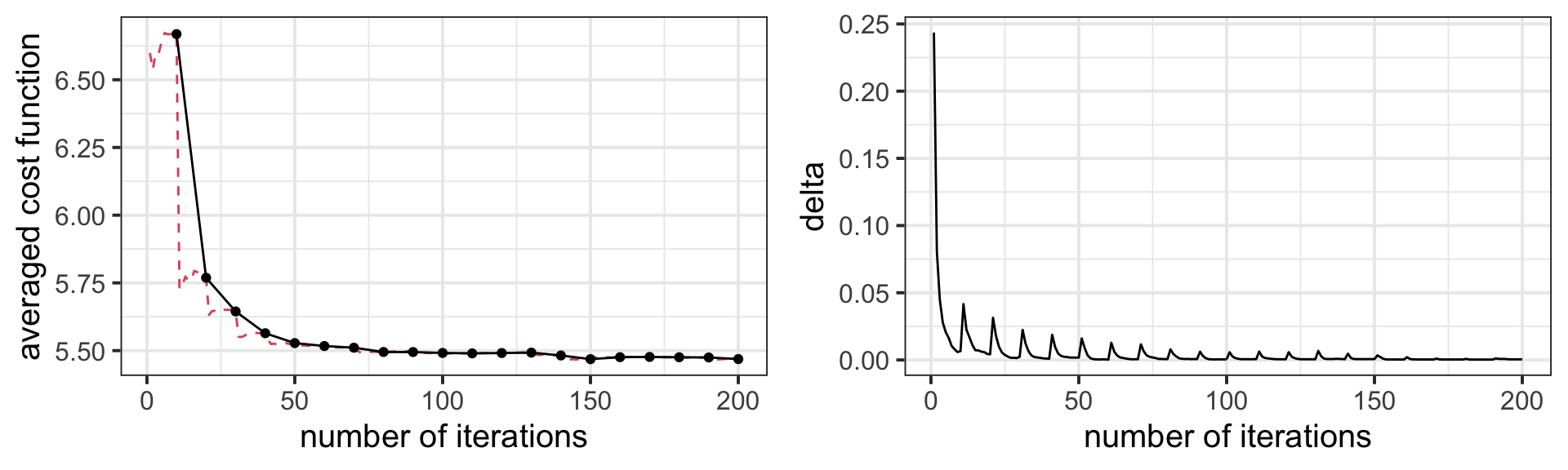}
    \vspace{-.25cm}
    \caption{\small Left: Average cost function at each iteration of the inner (red, dashed) and outer (black, solid) loop. Right: Relative change in estimated warping functions $\delta$.}
    \label{fig:algm_conv}
\end{figure}

The first quantity of interest is the average of the cost function in \eqref{eq:objfn} across all components $j\in [K]$. The second is the relative change in the estimated warping functions between iterations: $\delta(k) = \frac1{Kn}\sum_{i=1}^n \sum_{j=1}^K \|\hat\psi_{ij}^{(k)}-\hat\psi_{ij}^{(k-1)}\|^2,\ k=2,\dots$. The left panel in Figure \ref{fig:algm_conv} is a plot of the average cost function after each inner (red) and outer (black) loop update with respect to the iteration counter. It appears that the algorithm has converged after approximately $100$ iterations. The right panel in Figure \ref{fig:algm_conv} plots the value of $\delta$ with respect to the iteration counter. Overall, the value of $\delta$ decreases with jumps occurring after an update of the component templates. The magnitudes of these jumps gradually converge to 0 as the component templates also converge. 

\subsection{Comparison with regularization towards identity warping}

The main advantage of the proposed registration method is the regularization of estimated warping functions by use of the spatial penalty term. Regularization reduces the complexity and variance of the estimated phase component and guards against `overalignment' of function features that may be due to noise. In existing functional data analysis literature, most regularized registration methods use a regularization penalty that forces estimated warping functions to be close to the identity element, with respect to some prespecified distance on $\Gamma$ \citep{srivastava2016functional}, e.g., the squared extrinsic Fisher-Rao distance given by $\|\psi-1\|^2$, where $\psi = Q(\gamma)$ and $1=Q(\gamma_{id})$; see Section \ref{sec:pairwiseunivariate}. However, in the case of multivariate functional data wherein phase is spatially dependent, the correlation structure in cross-component phase enables us to regularize estimated warping functions toward data-driven targets rather than the fixed identity warping function. To assess this benefit, we compare the estimated warping functions generated using the proposed approach to a regularized version of componentwise registration as specified by \eqref{eq:objfn1D}. 

\begin{figure}[!t]
    \centering
    \begin{tabular}{cc}
    (a)&(b)\\
    \includegraphics[scale=0.11]{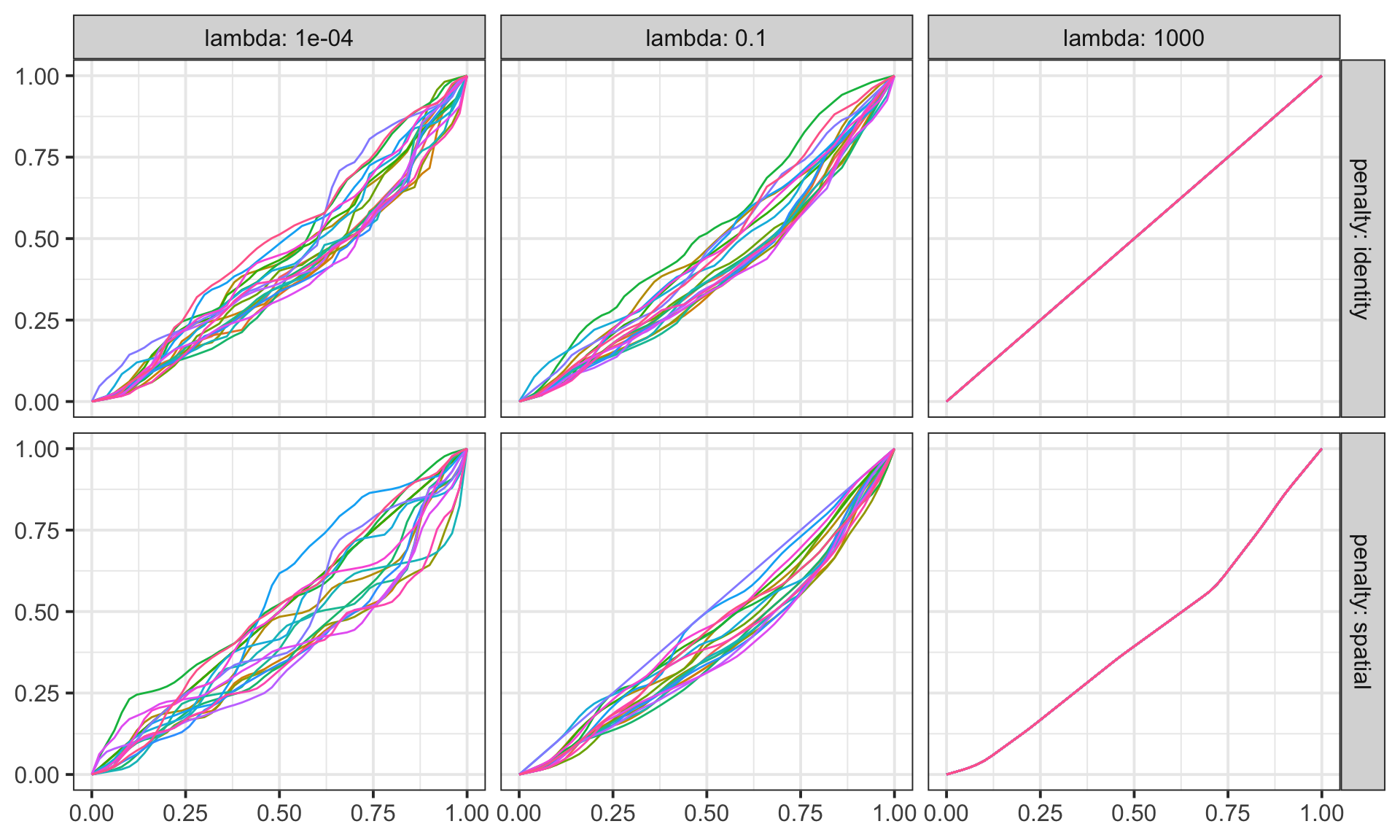}&\includegraphics[scale=0.11]{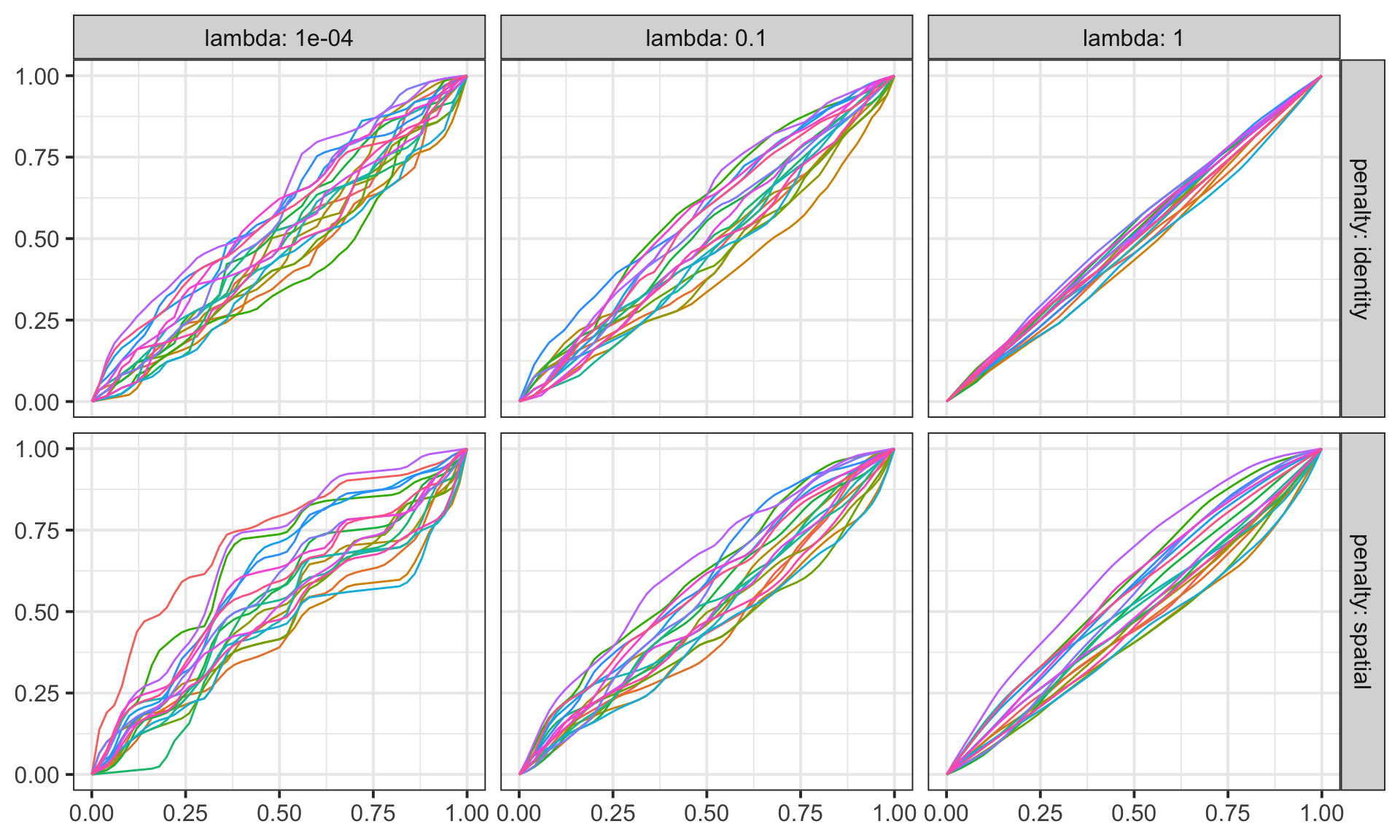}
    \end{tabular}
    \caption{\small Estimated warping functions for (a) all components of a fixed observation $i$ and (b) a fixed component at location $\mathbf s_j$ of all observations, under Simulation setting 1 for different values of the regularization parameter $\lambda$. Top row: Independent componentwise regularized registration. Bottom row: Proposed approach.}
    \label{fig:penalty_cc}
\end{figure}

Figure \ref{fig:penalty_cc}(a) presents estimated warping functions for all $j \in [K]$ components of the same observation $i$ in a randomly selected simulation run under Simulation setting 1. We consider three different values for the regularization parameter: $\lambda = 1\times 10^{-4}$, $ 1\times 10^{-1}$, and $1\times 10^{3}$. The results of regularized componentwise registration are given in the top row while the results generated by the proposed approach are shown in the bottom row. 

The magnitude and roughness of the cross-component phase estimated using regularized componentwise registration are penalized toward the identity warping;  in regularized componentwise registration, when $\lambda=1000$, the estimated warping functions all converge to the same identity element. On the other hand, the proposed approach allows estimated warping functions to converge to a more flexible target (see bottom right panel of Figure \ref{fig:penalty_cc}(a)) when $\lambda$ increases. This is because the \emph{kriging estimate $\tilde \psi_{ij}$ in the penalty can vary from observation to observation}. This phenomenon is also observed in Figure \ref{fig:penalty_cc}(b) where we fix a component $j$ and display the estimated cross-observation phases. The regularized componentwise registration forces all warping functions towards the identity when $\lambda$ is greater than 1. These results imply very small cross-observation phase variation, which is not true for the simulated data. On the other hand, the proposed penalized registration captures the cross-observation phase variation even if $\lambda$ is fairly large. Based on these findings, \emph{we see that the proposed approach reduces the discrepancy of estimated warping functions for components in each observation and preserves the magnitude of estimated cross-observation phases}. The proposed approach using the spatial penalty results in warping functions with appropriate smoothness and magnitude, but does not force them toward identity regardless of cross-observation phase variation.

\section{Real data analysis}
\label{sec:realdata}

We apply the proposed penalized registration method to two multivariate functional datasets. The first consists of multi-trial EEG signals for alcoholism patients discussed earlier, and the second considers annual ozone concentration functions observed at different locations in a small area in northern California. In both cases, we use 4-fold cross-validation to select a value for $\lambda$ in the proposed registration approach.

\subsection{Electroencephalogram data}
\label{sec:EEGanalysis}

We analyze an EEG dataset arising from a study that examined EEG correlates of genetic predisposition to alcoholism \citep{bache2013uci}. The study resulted in EEG measurements recorded at 64 electrodes placed on the subjects' scalps, with a sampling rate of 256 Hz (3.9-msec epoch) for 1 second. Each subject was exposed to a stimulus in each trial and completed multiple trials where different stimuli were shown. The study used standard 64-channel electrode placements as defined by the American Electroencephalographic Association. In our analysis, we use EEG signals from 50 trials for one subject, and treat each trial as an independent multivariate functional observation. Prior to analysis, each EEG signal was smoothed using smoothing splines with a small parameter value of $1\times 10^{-5}$. To compare the performance of different methods, we register the 50 trials using the componentwise, universal and proposed registration approaches. After registration, we average the aligned signals recorded by each electrode across trials to estimate the mean event related potentials.
 
\begin{figure}[!t]
    \centering
    \includegraphics[width =1\textwidth]{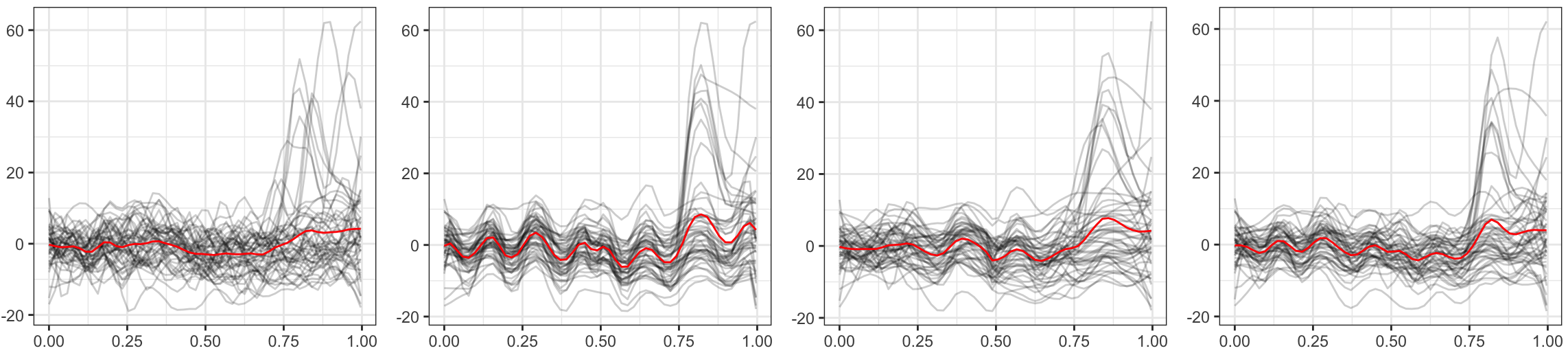}
    \vspace{-.75cm}
    \caption{\small Registration and estimate $\hat \mu_{\mathbf s_j}$ of template component $\mu_{\mathbf s_j}$ at electrode AFZ (location $\mathbf s_j$). Panel 1: Simulated data in grey with cross-sectional average without registration in red. Panels 2-4: Aligned functions in grey with estimated average signal in red, generated using componentwise registration, universal registration, and the proposed method, respectively.}
    \label{fig:alcohol_example}
\end{figure}

For easy comparison across the three methods, we display the registration results for electrode AFZ in Figure \ref{fig:alcohol_example}. As in the simulations, we also display the result of averaging across trials when no registration is applied to the data as a baseline (Panel 1). The electrode AFZ was chosen as an illustration since many of the EEG signals recorded there contain a pronounced activation peak between $t=0.75$ and $t=1$. We observe similar results in this real EEG data example, based on the different registration methods, as in the simulation studies. In Panel 1, when no alignment is applied, the pronounced activation peak is almost completely nonexistent in the average signal shown in red; thus, registration of the data is necessary prior to averaging. In Panel 2, the componentwise registration method tends to `overalign' many of the relatively small modes present in the signals that are likely due to noise. This results in an average that has many small fluctuations that are indistinguishable from the pronounced activation peak. In Panels 3 and 4, the universal and proposed methods are effective in capturing the large peak in the respective average signals and tend to produce fewer small fluctuations prior to $t=0.75$. Essentially, these two methods mitigate the contribution of noise and result in average signals that reflect the most prominent feature in the given data. The magnitude of the pronounced activation peak is similar in the two averages, but it is stretched over a longer part of the domain in the average produced using the universal method. 

\begin{figure}[!t]
    \centering
    \includegraphics[width =1\textwidth]{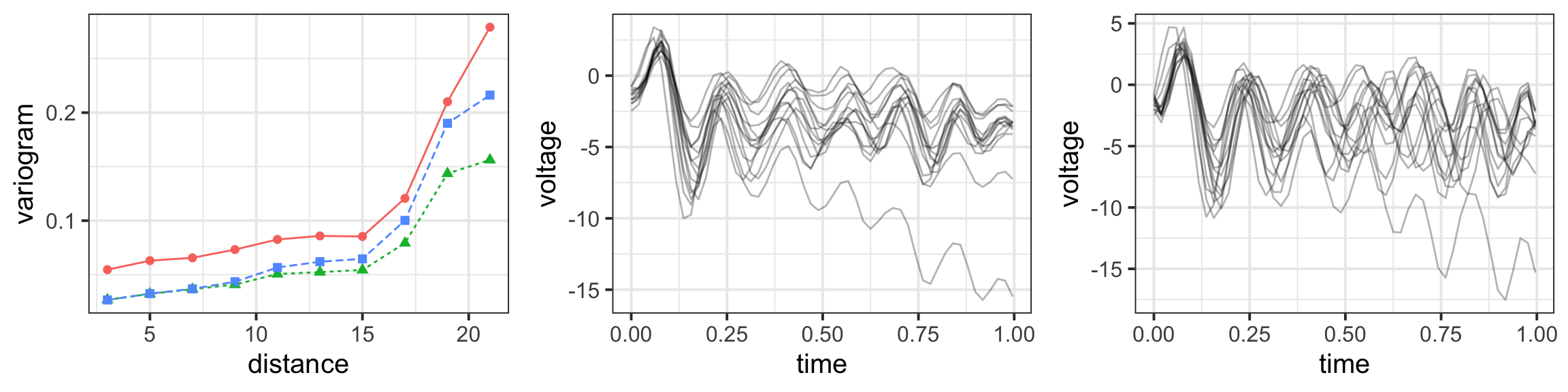}
    \vspace{-.75cm}
    \caption{\small Empirical trace-variograms $\hat V$ (left) computed using estimated average signals $\{\hat \mu_{\mathbf s_j}, j \in [64]\}$ obtained using three registration methods: componentwise (red, solid), universal (blue, dashed), and proposed (green, dotted). Estimated average EEG signals $\{\hat \mu_{\mathbf s_1},\ldots, \hat \mu_{\mathbf s_{14}}\}$ at 14 (out of the total 64) electrodes placed in an area related to the parietal lobe, computed using the proposed (middle) and componentwise (right) registration approaches.}
    \label{fig:alcohol_variogram}
\end{figure}

In EEG data analysis, it is often of interest to study the relationship between average signals (across trials) at different electrode locations after registration. To show the benefits of the proposed approach in downstream EEG data analysis tasks, we explore the spatial correlation between estimated average EEG signals using the empirical trace-variogram \citep{giraldo2011ordinary}, defined as 
\begin{align}\label{eq:emp_vargm_alcohol}
\hat V(h)=\frac{1}{2|N_\epsilon(h)|} \sum \limits_{a,b\in N_\epsilon(h)} \|\hat \mu_{s_a}-\hat \mu_{s_b}\|^2,
\end{align}
where $N_\epsilon (h)= \{( \mathbf  s_a, \mathbf  s_b):\| \mathbf  s_a- \mathbf  s_b\|\in (h-\epsilon,h+\epsilon)\}$ for a small $\epsilon >0$ and $\hat \mu_{s_a}$ denotes the estimated average signal for electrode located at $\mathbf  s_a,\ a = 1,\ldots,64$. The empirical trace-variograms, computed based on average signals generated by the three different registration methods, are shown in the left panel of Figure \ref{fig:alcohol_variogram} (componentwise in red, universal in blue, proposed in green). The variograms suggest very similar spatial correlation patterns between the average signals, but the proposed approach produces averages with smallest spatial variation. The universal approach results in more spatial variation due to its very restrictive assumption of common phase across all channels. The magnitude of spatial variation in this case is similar to the proposed method at small spatial distances. This is intuitive since cross-component phase variation at nearby electrodes is very small and the common phase assumption is reasonable. However, as the spatial distance increases, this assumption becomes unrealistic.

The componentwise method conducts separate alignment of EEG signals at each electrode, resulting in large phase variation across electrode locations. On the other hand, the proposed method avoids this issue by accounting for the spatial phase correlation across electrodes through the regularization penalty. To demonstrate this, we consider the 14 electrodes in the area corresponding to the parietal lobe. Since all of the 14 electrodes are related to the same brain region, we should expect very little phase variation in the resulting average signals. The corresponding estimated averages are shown in the middle panel in Figure \ref{fig:alcohol_variogram} for the proposed method and the right panel in the same figure for the componentwise method. In the middle panel, there is very little phase variation across the estimated averages, as expected. However, in the right panel, it is easy to see that considerable phase variation remains. 

\subsection{Ozone concentration data}

Ground-level ozone is a harmful pollutant that is monitored closely by the Environmental Protection Agency. The temporal trend of daily ozone concentration varies from year to year due to variations in various environmental conditions, including the weather. To explore average ozone concentration patterns at a certain location, we desire to first eliminate phase variation that exists across different observation years using a registration procedure. At the same time, we must account for cross-component phase variation across different locations within the same year that is due to different temporal patterns of pollutant spread. Thus, we view multi-year ozone concentration functions, observed at different spatial locations, as multivariate functional data with two different sources of phase variation; here, univariate ozone concentration functions observed at different spatial locations are treated as components of a full observation corresponding to a single year. In this analysis, we focus on a small area in northern California ($35^{\circ}\sim 39^{\circ}$ N, $120\sim 123^{\circ}$ W) with $K=16$ observation stations (components). Each station recorded daily average ozone concentration (in parts per million) from year 2000 to year 2019 (sample size $n=20$). The data is publicly available on the air data website\footnote{\url{https://www.epa.gov/outdoor-air-quality-data}} of the United States Environmental Protection Agency. As in the previous real data analysis example that considered EEG data, we compare the performance of three registration procedures in this context: componentwise, universal and proposed.

\begin{figure}[!t]
    \centering
    \includegraphics[width =1\textwidth]{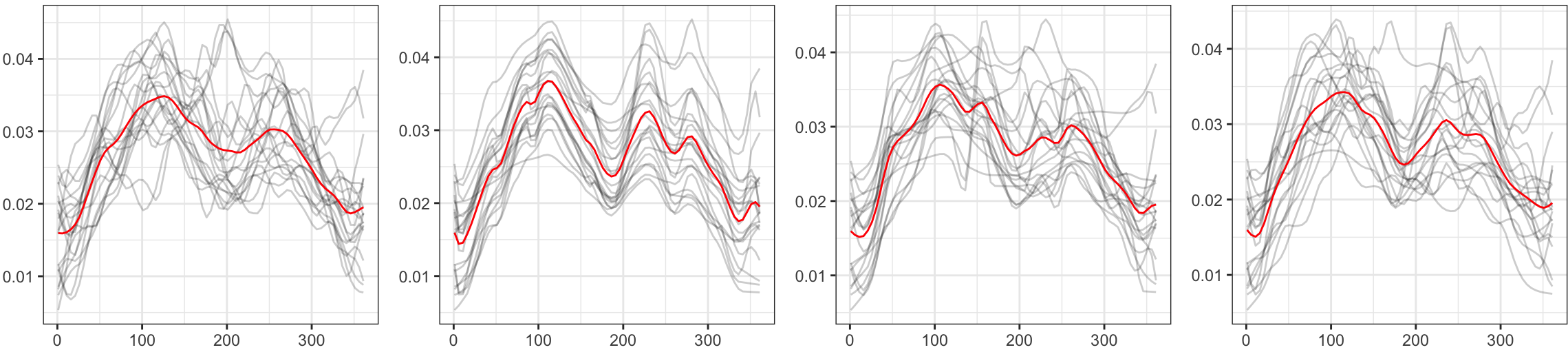}
    \vspace{-.75cm}
    \caption{\small Registration and estimate $\hat \mu_{\mathbf s_j}$ of template component $\mu_{\mathbf s_j}$ where $\mathbf s_j$ is  Alameda County (37.8 N, 122.3 W). Panel 1: Observed data in grey with cross-sectional average without registration in red. Panels 2-4: Aligned functions in grey with estimated average signal in red, generated using componentwise registration, universal registration, and the proposed method.}
    \label{fig:ozone_example}
\end{figure}

We display the registration results for a single location, corresponding to Alameda County (37.8 N, 122.3 W), in Figure \ref{fig:ozone_example}. We also display the result of averaging across trials when no registration is applied to the data as a baseline (Panel 1). We make several interesting observations based on these results. First, the proposed method yields an estimate of the average ozone concentration function (Panel 4) that contains clearer patterns than the average computed without alignment (Panel 1), e.g., the steep reduction in ozone concentration around day 200. The proposed approach produces an average that has very similar patterns to those computed using the componentwise and universal registration methods, but is smoother overall. The most noticeable difference occurs in early autumn where the proposed approach results in an average that has a single mode, while the other two methods result in averages with two smaller peaks. A possible reason for this phenomenon is the large regularization parameter value chosen via cross-validation; this restricts the complexity of estimated cross-component warping functions and forces the registration procedure to overlook small features that are potentially induced by noise. 

\begin{figure}[!t]
    \centering
    \includegraphics[width =1\textwidth]{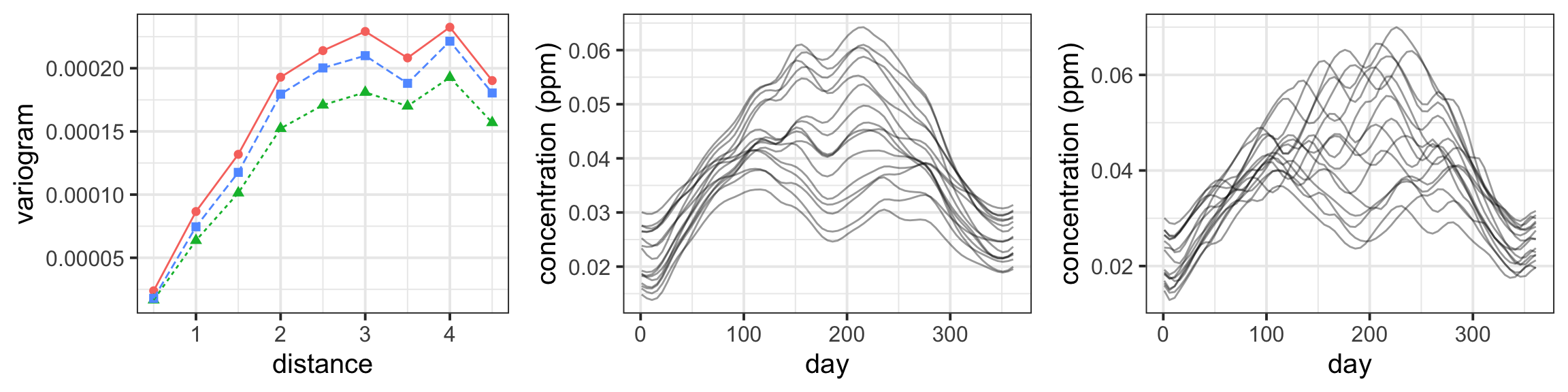}
    \vspace{-.75cm}
    \caption{\small Empirical trace-variograms $\hat V$ (left) computed using estimated average ozone concentration functions $\{\hat \mu_{\mathbf s_j},\ j \in [16]\}$ at 16 locations in a small area in northern California obtained using three registration methods: componentwise (red, solid), universal (blue, dashed), and proposed (green, dotted). Estimated average ozone concentration functions $\{\hat \mu_{\mathbf s_j},\ j \in [16]\}$ computed using the proposed (middle) and componentwise (right) registration approaches.}
    \label{fig:ozone_variogram}
\end{figure}

As in the EEG data example, we further analyze the estimated average ozone concentration functions at the 16 spatial locations. Modeling of spatial correlation based on averages computed after registration is also of particular interest in this application. In the left panel of Figure \ref{fig:ozone_variogram}, we display the empirical trace-variograms, computed using \eqref{eq:emp_vargm_alcohol} after estimating the average ozone concentration functions using the componentwise (red), universal (blue) and proposed (green) approaches. The conclusions here are similar to those reached in the EEG data example. The proposed method yields componentwise averages that have smallest spatial variation. The variation gap between the universal and proposed methods appears bigger in this case, even for small spatial distances. We further display the componentwise estimated average ozone concentration functions, for each of the 16 locations, generated using the proposed and componentwise registration methods in the middle and right panels of Figure \ref{fig:ozone_variogram}, respectively. It is obvious that the proposed approach (middle) results in estimated average functions that exhibit much less cross-component phase variation than the average functions generated using the componentwise registration method (right). Furthermore, most averages in the middle panel have two ozone concentration maxima around April/May and August, and a single ozone concentration minimum around June/July. In the right panel, there is much more variation in the number of extrema in the estimated averages as well as their timing.

\section{Discussion}
The proposed penalized registration procedure can in principle be used to align multivariate functional data with component functions that are correlated in other ways. For example, when component functions are temporally correlated, we can replace the kriging estimate $\tilde \psi_{ij}$ in the penalty term of the objective function \eqref{eq:objfn_pen} with a temporal interpolant or temporally weighted estimate (e.g., moving time-window average). There is much to be done in this direction, and the proposed method represents a promising initial foray. 

The focus in this paper was restricted to multivariate functional data wherein \emph{only} the cross-component phases are spatially correlated. The objective function \eqref{eq:objfn_pen} does not quantify and incorporate any spatial correlation between the cross-component amplitudes, and this may well be of interest in certain applications. Extension of the algorithm to accommodate such a requirement is possible using the \emph{amplitude trace-variogram} proposed by \cite{guo2020variograms}, in addition to the phase trace-variogram used here. This presents a fruitful line for future work. 

Importance of the elastic metric and its invariance to time warping, which is made practically useful through the square-root slope function (SRSF) representation, cannot be overstated. The invariance property is used to great benefit in both the objective function and the novel spatially-informed penalty term through the kriging estimate $\tilde \psi_{ij}$, and enables us to disregard estimating the cross-observation phases $\{\alpha_i\}$. 

There is potential for improvement in reducing the computational burden when implementing the registration procedure when the number $n$ of functional observations with $K$ components are both large. Registration in the elastic framework uses the the dynamic programming algorithm since the class of warping functions $\Gamma$ is infinite-dimensional and unconstrained \citep{srivastava2016functional}. Restricting attention to a smaller parametric class (e.g., parameterised class of distribution or quantile functions on $[0,1]$) will reduce computing time considerably, but at the cost of flexibility in registration.

\noindent\textbf{Acknowledgements:} We acknowledge funding from the National Science Foundation (DMS-2015374 to KB; CCF-1740761, DMS-2015226, CCF-1839252 to SK), the National Institutes of Health (R37-CA214955 to SK and
KB) and the Engineering and Physical Sciences Research Council (EP/V048104/1 to KB).
  \bibliography{eeg_align_ref.bib}{} 

\begin{thebibliography}{17}
\expandafter\ifx\csname natexlab\endcsname\relax\def\natexlab#1{#1}\fi

\bibitem[{Bache \& Lichman(2013)}]{bache2013uci}
\textsc{Bache, K.} \& \textsc{Lichman, M.} (2013).
\newblock {UCI} machine learning repository.
\newblock \url{https://archive.ics.uci.edu/ml/datasets/EEG+Database}.

\bibitem[{Carroll \& M\"{u}ller(2021)}]{carroll2021}
\textsc{Carroll, C.} \& \textsc{M\"{u}ller, H.-G.} (2021).
\newblock Latent transport models for multivariate functional data.
\newblock \textit{arXiv:2107.05730} .

\bibitem[{Carroll et~al.(2021)Carroll, M{\"u}ller \& Kneip}]{carroll2020cross}
\textsc{Carroll, C.}, \textsc{M{\"u}ller, H.-G.} \& \textsc{Kneip, A.} (2021).
\newblock Cross-component registration for multivariate functional data, with
  application to growth curves.
\newblock \textit{Biometrics} \textbf{77}, 839--851.

\bibitem[{Giraldo et~al.(2011)Giraldo, Delicado \& Mateu}]{giraldo2011ordinary}
\textsc{Giraldo, R.}, \textsc{Delicado, P.} \& \textsc{Mateu, J.} (2011).
\newblock Ordinary kriging for function-valued spatial data.
\newblock \textit{Environmental and Ecological Statistics} \textbf{18},
  411--426.

\bibitem[{Guo et~al.(2022)Guo, Kurtek \& Bharath}]{guo2020variograms}
\textsc{Guo, X.}, \textsc{Kurtek, S.} \& \textsc{Bharath, K.} (2022).
\newblock Variograms for kriging and clustering of spatial functional data with
  phase variation.
\newblock \textit{Spatial Statistics} , 100687.

\bibitem[{Kurtek et~al.(2012)Kurtek, Srivastava, Klassen \&
  Ding}]{kurtek2012statistical}
\textsc{Kurtek, S.}, \textsc{Srivastava, A.}, \textsc{Klassen, E.} \&
  \textsc{Ding, Z.} (2012).
\newblock Statistical modeling of curves using shapes and related features.
\newblock \textit{Journal of the American Statistical Association}
  \textbf{107}, 1152--1165.

\bibitem[{Makeig et~al.(2007)Makeig, Onton, Sejnowski \&
  Poizner}]{makeig2007prospects}
\textsc{Makeig, S.}, \textsc{Onton, J.}, \textsc{Sejnowski, T.} \&
  \textsc{Poizner, H.} (2007).
\newblock Prospects for mobile, high-definition brain imaging: {EEG} spectral
  modulations during {3-D} reaching.
\newblock \textit{Human Brain Mapping} .

\bibitem[{Marron et~al.(2015)Marron, Ramsay, Sangalli \&
  Srivastava}]{marron2015functional}
\textsc{Marron, J.~S.}, \textsc{Ramsay, J.~O.}, \textsc{Sangalli, L.~M.} \&
  \textsc{Srivastava, A.} (2015).
\newblock Functional data analysis of amplitude and phase variation.
\newblock \textit{Statistical Science} \textbf{30}, 468--484.

\bibitem[{Olsen et~al.(2016)Olsen, Markussen \& Raket}]{olsen2016simultaneous}
\textsc{Olsen, N.}, \textsc{Markussen, B.} \& \textsc{Raket, L.~L.} (2016).
\newblock Simultaneous inference for misaligned multivariate functional data.
\newblock \textit{Journal of the Royal Statistical Society: Series C}
  \textbf{67}, 1147--1176.

\bibitem[{Park \& Ahn(2017)}]{park2017clustering}
\textsc{Park, J.} \& \textsc{Ahn, J.} (2017).
\newblock Clustering multivariate functional data with phase variation.
\newblock \textit{Biometrics} \textbf{73}, 324--333.

\bibitem[{Srivastava \& Klassen(2016)}]{srivastava2016functional}
\textsc{Srivastava, A.} \& \textsc{Klassen, E.~P.} (2016).
\newblock \textit{Functional and Shape Data Analysis}.
\newblock Springer.

\bibitem[{Srivastava et~al.(2011)Srivastava, Wu, Kurtek, Klassen \&
  Marron}]{srivastava2011registration}
\textsc{Srivastava, A.}, \textsc{Wu, W.}, \textsc{Kurtek, S.}, \textsc{Klassen,
  E.} \& \textsc{Marron, J.~S.} (2011).
\newblock Registration of functional data using {F}isher-{R}ao metric.
\newblock \textit{arXiv:1103.3817} .

\bibitem[{Stam et~al.(2007)Stam, Nolte \& Daffertshofer}]{stam2007phase}
\textsc{Stam, C.~J.}, \textsc{Nolte, G.} \& \textsc{Daffertshofer, A.} (2007).
\newblock Phase lag index: assessment of functional connectivity from multi
  channel {EEG} and {MEG} with diminished bias from common sources.
\newblock \textit{Human Brain Mapping} \textbf{28}, 1178--1193.

\bibitem[{Sur \& Sinha(2009)}]{sur2009event}
\textsc{Sur, S.} \& \textsc{Sinha, V.~K.} (2009).
\newblock Event-related potential: An overview.
\newblock \textit{Industrial Psychiatry Journal} \textbf{18}, 70--73.

\bibitem[{Tsai et~al.(2014)Tsai, Jung, Chien, Savostyanov \&
  Makeig}]{tsai2014cortical}
\textsc{Tsai, A.~C.}, \textsc{Jung, T.-P.}, \textsc{Chien, V.~S.},
  \textsc{Savostyanov, A.~N.} \& \textsc{Makeig, S.} (2014).
\newblock Cortical surface alignment in multi-subject spatiotemporal
  independent {EEG} source imaging.
\newblock \textit{NeuroImage} \textbf{87}, 297--310.

\bibitem[{Wang et~al.(2001)Wang, Begleiter \& Porjesz}]{wang2001warp}
\textsc{Wang, K.}, \textsc{Begleiter, H.} \& \textsc{Porjesz, B.} (2001).
\newblock Warp-averaging event-related potentials.
\newblock \textit{Clinical Neurophysiology} \textbf{112}, 1917--1924.

\bibitem[{Zhao et~al.(2020)Zhao, Xu, Li \& Wu}]{zhao2020modeling}
\textsc{Zhao, W.}, \textsc{Xu, Z.}, \textsc{Li, W.} \& \textsc{Wu, W.} (2020).
\newblock Modeling and analyzing neural signals with phase variability using
  {F}isher-{R}ao registration.
\newblock \textit{Journal of Neuroscience Methods} \textbf{346}, 108954.

\end{thebibliography}

\newpage
\appendix


\end{document}